\documentclass[prl,preprintnumbers,twocolumn,floatfix,nofootinbib,a4paper,superscriptaddress]{revtex4-2}

\newif\ifprintauthors
\printauthorstrue

\newif\ifprintsupplement
\printsupplementtrue
\usepackage{xcolor}
\usepackage{multirow}
\usepackage{tabularx}
\usepackage{graphicx}
\usepackage{dcolumn}
\usepackage{bm}
\usepackage{acronym}
\usepackage{hyperref}
\usepackage{threeparttable}
\usepackage{amsmath}
\usepackage{lineno}
\usepackage{xr}
\usepackage{xspace}
\usepackage{amsfonts}
\usepackage{booktabs}
\usepackage{orcidlink}
\usepackage{etoolbox}

\ifprintsupplement
\newcommand{\citesupp}{}
\else
\newcommand{\citesupp}{ \cite{supplement}}
\fi

\newcommand{\aoneminus}[1]{\IfEqCase{#1}{{GW250114_NRSUR}{0.09}{GW250114_SEOB}{0.13}{GW250114_XO4a}{0.13}{GW250114_Phenom}{0.11}}}
\newcommand{\aonemed}[1]{\IfEqCase{#1}{{GW250114_NRSUR}{0.09}{GW250114_SEOB}{0.13}{GW250114_XO4a}{0.13}{GW250114_Phenom}{0.11}}}
\newcommand{\aoneplus}[1]{\IfEqCase{#1}{{GW250114_NRSUR}{0.15}{GW250114_SEOB}{0.19}{GW250114_XO4a}{0.17}{GW250114_Phenom}{0.21}}}
\newcommand{\aoneninetypercent}[1]{\IfEqCase{#1}{{GW250114_NRSUR}{0.24}{GW250114_SEOB}{0.32}{GW250114_XO4a}{0.30}{GW250114_Phenom}{0.32}}}
\newcommand{\aonetenpercent}[1]{\IfEqCase{#1}{{GW250114_NRSUR}{0.02}{GW250114_SEOB}{0.03}{GW250114_XO4a}{0.03}{GW250114_Phenom}{0.02}}}
\newcommand{\atwominus}[1]{\IfEqCase{#1}{{GW250114_NRSUR}{0.10}{GW250114_SEOB}{0.14}{GW250114_XO4a}{0.16}{GW250114_Phenom}{0.13}}}
\newcommand{\atwomed}[1]{\IfEqCase{#1}{{GW250114_NRSUR}{0.10}{GW250114_SEOB}{0.14}{GW250114_XO4a}{0.16}{GW250114_Phenom}{0.13}}}
\newcommand{\atwoplus}[1]{\IfEqCase{#1}{{GW250114_NRSUR}{0.16}{GW250114_SEOB}{0.20}{GW250114_XO4a}{0.19}{GW250114_Phenom}{0.23}}}
\newcommand{\atwoninetypercent}[1]{\IfEqCase{#1}{{GW250114_NRSUR}{0.26}{GW250114_SEOB}{0.34}{GW250114_XO4a}{0.35}{GW250114_Phenom}{0.36}}}
\newcommand{\atwotenpercent}[1]{\IfEqCase{#1}{{GW250114_NRSUR}{0.02}{GW250114_SEOB}{0.03}{GW250114_XO4a}{0.03}{GW250114_Phenom}{0.02}}}
\newcommand{\chieffminus}[1]{\IfEqCase{#1}{{GW250114_NRSUR}{0.04}{GW250114_SEOB}{0.05}{GW250114_XO4a}{0.05}{GW250114_Phenom}{0.05}}}
\newcommand{\chieffmed}[1]{\IfEqCase{#1}{{GW250114_NRSUR}{-0.03}{GW250114_SEOB}{-0.05}{GW250114_XO4a}{-0.07}{GW250114_Phenom}{-0.03}}}
\newcommand{\chieffplus}[1]{\IfEqCase{#1}{{GW250114_NRSUR}{0.03}{GW250114_SEOB}{0.04}{GW250114_XO4a}{0.06}{GW250114_Phenom}{0.04}}}
\newcommand{\chieffninetypercent}[1]{\IfEqCase{#1}{{GW250114_NRSUR}{-0.01}{GW250114_SEOB}{-0.02}{GW250114_XO4a}{-0.03}{GW250114_Phenom}{0.00}}}
\newcommand{\chiefftenpercent}[1]{\IfEqCase{#1}{{GW250114_NRSUR}{-0.07}{GW250114_SEOB}{-0.09}{GW250114_XO4a}{-0.11}{GW250114_Phenom}{-0.07}}}
\newcommand{\chipminus}[1]{\IfEqCase{#1}{{GW250114_NRSUR}{0.11}{GW250114_SEOB}{0.15}{GW250114_XO4a}{0.13}{GW250114_Phenom}{0.15}}}
\newcommand{\chipmed}[1]{\IfEqCase{#1}{{GW250114_NRSUR}{0.11}{GW250114_SEOB}{0.15}{GW250114_XO4a}{0.16}{GW250114_Phenom}{0.15}}}
\newcommand{\chipplus}[1]{\IfEqCase{#1}{{GW250114_NRSUR}{0.15}{GW250114_SEOB}{0.19}{GW250114_XO4a}{0.16}{GW250114_Phenom}{0.21}}}
\newcommand{\chipninetypercent}[1]{\IfEqCase{#1}{{GW250114_NRSUR}{0.27}{GW250114_SEOB}{0.33}{GW250114_XO4a}{0.31}{GW250114_Phenom}{0.36}}}
\newcommand{\chiptenpercent}[1]{\IfEqCase{#1}{{GW250114_NRSUR}{0.04}{GW250114_SEOB}{0.05}{GW250114_XO4a}{0.07}{GW250114_Phenom}{0.05}}}
\newcommand{\chirpmassminus}[1]{\IfEqCase{#1}{{GW250114_NRSUR}{0.4}{GW250114_SEOB}{0.5}{GW250114_XO4a}{0.5}{GW250114_Phenom}{0.5}}}
\newcommand{\chirpmassmed}[1]{\IfEqCase{#1}{{GW250114_NRSUR}{31.1}{GW250114_SEOB}{30.9}{GW250114_XO4a}{30.8}{GW250114_Phenom}{31.2}}}
\newcommand{\chirpmassplus}[1]{\IfEqCase{#1}{{GW250114_NRSUR}{0.4}{GW250114_SEOB}{0.5}{GW250114_XO4a}{0.6}{GW250114_Phenom}{0.4}}}
\newcommand{\chirpmassninetypercent}[1]{\IfEqCase{#1}{{GW250114_NRSUR}{31.4}{GW250114_SEOB}{31.3}{GW250114_XO4a}{31.3}{GW250114_Phenom}{31.5}}}
\newcommand{\chirpmasstenpercent}[1]{\IfEqCase{#1}{{GW250114_NRSUR}{30.8}{GW250114_SEOB}{30.6}{GW250114_XO4a}{30.4}{GW250114_Phenom}{30.8}}}
\newcommand{\chirpmasssourceminus}[1]{\IfEqCase{#1}{{GW250114_NRSUR}{0.5}{GW250114_SEOB}{0.5}{GW250114_XO4a}{0.6}{GW250114_Phenom}{0.5}}}
\newcommand{\chirpmasssourcemed}[1]{\IfEqCase{#1}{{GW250114_NRSUR}{28.6}{GW250114_SEOB}{28.6}{GW250114_XO4a}{28.5}{GW250114_Phenom}{28.7}}}
\newcommand{\chirpmasssourceplus}[1]{\IfEqCase{#1}{{GW250114_NRSUR}{0.5}{GW250114_SEOB}{0.5}{GW250114_XO4a}{0.6}{GW250114_Phenom}{0.5}}}
\newcommand{\chirpmasssourceninetypercent}[1]{\IfEqCase{#1}{{GW250114_NRSUR}{29.0}{GW250114_SEOB}{29.0}{GW250114_XO4a}{29.0}{GW250114_Phenom}{29.1}}}
\newcommand{\chirpmasssourcetenpercent}[1]{\IfEqCase{#1}{{GW250114_NRSUR}{28.2}{GW250114_SEOB}{28.2}{GW250114_XO4a}{28.1}{GW250114_Phenom}{28.3}}}
\newcommand{\finalmassminus}[1]{\IfEqCase{#1}{{GW250114_NRSUR}{0.9}{GW250114_SEOB}{1.0}{GW250114_XO4a}{1.1}{GW250114_Phenom}{1.0}}}
\newcommand{\finalmassmed}[1]{\IfEqCase{#1}{{GW250114_NRSUR}{68.1}{GW250114_SEOB}{67.8}{GW250114_XO4a}{67.6}{GW250114_Phenom}{68.2}}}
\newcommand{\finalmassplus}[1]{\IfEqCase{#1}{{GW250114_NRSUR}{0.8}{GW250114_SEOB}{1.0}{GW250114_XO4a}{1.1}{GW250114_Phenom}{0.9}}}
\newcommand{\finalmassninetypercent}[1]{\IfEqCase{#1}{{GW250114_NRSUR}{68.7}{GW250114_SEOB}{68.6}{GW250114_XO4a}{68.4}{GW250114_Phenom}{68.9}}}
\newcommand{\finalmasstenpercent}[1]{\IfEqCase{#1}{{GW250114_NRSUR}{67.4}{GW250114_SEOB}{67.0}{GW250114_XO4a}{66.7}{GW250114_Phenom}{67.5}}}
\newcommand{\finalmasssourceminus}[1]{\IfEqCase{#1}{{GW250114_NRSUR}{1.1}{GW250114_SEOB}{1.0}{GW250114_XO4a}{1.1}{GW250114_Phenom}{1.0}}}
\newcommand{\finalmasssourcemed}[1]{\IfEqCase{#1}{{GW250114_NRSUR}{62.7}{GW250114_SEOB}{62.6}{GW250114_XO4a}{62.5}{GW250114_Phenom}{62.9}}}
\newcommand{\finalmasssourceplus}[1]{\IfEqCase{#1}{{GW250114_NRSUR}{1.0}{GW250114_SEOB}{1.0}{GW250114_XO4a}{1.2}{GW250114_Phenom}{1.1}}}
\newcommand{\finalmasssourceninetypercent}[1]{\IfEqCase{#1}{{GW250114_NRSUR}{63.5}{GW250114_SEOB}{63.4}{GW250114_XO4a}{63.4}{GW250114_Phenom}{63.7}}}
\newcommand{\finalmasssourcetenpercent}[1]{\IfEqCase{#1}{{GW250114_NRSUR}{61.9}{GW250114_SEOB}{61.8}{GW250114_XO4a}{61.6}{GW250114_Phenom}{62.1}}}
\newcommand{\finalspinminus}[1]{\IfEqCase{#1}{{GW250114_NRSUR}{0.01}{GW250114_SEOB}{0.01}{GW250114_XO4a}{0.01}{GW250114_Phenom}{0.01}}}
\newcommand{\finalspinmed}[1]{\IfEqCase{#1}{{GW250114_NRSUR}{0.68}{GW250114_SEOB}{0.67}{GW250114_XO4a}{0.67}{GW250114_Phenom}{0.68}}}
\newcommand{\finalspinplus}[1]{\IfEqCase{#1}{{GW250114_NRSUR}{0.01}{GW250114_SEOB}{0.01}{GW250114_XO4a}{0.02}{GW250114_Phenom}{0.01}}}
\newcommand{\finalspinninetypercent}[1]{\IfEqCase{#1}{{GW250114_NRSUR}{0.68}{GW250114_SEOB}{0.68}{GW250114_XO4a}{0.68}{GW250114_Phenom}{0.69}}}
\newcommand{\finalspintenpercent}[1]{\IfEqCase{#1}{{GW250114_NRSUR}{0.67}{GW250114_SEOB}{0.66}{GW250114_XO4a}{0.65}{GW250114_Phenom}{0.67}}}
\newcommand{\luminositydistanceminus}[1]{\IfEqCase{#1}{{GW250114_NRSUR}{70}{GW250114_SEOB}{69}{GW250114_XO4a}{72}{GW250114_Phenom}{72}}}
\newcommand{\luminositydistancemed}[1]{\IfEqCase{#1}{{GW250114_NRSUR}{403}{GW250114_SEOB}{385}{GW250114_XO4a}{381}{GW250114_Phenom}{399}}}
\newcommand{\luminositydistanceplus}[1]{\IfEqCase{#1}{{GW250114_NRSUR}{74}{GW250114_SEOB}{75}{GW250114_XO4a}{77}{GW250114_Phenom}{79}}}
\newcommand{\luminositydistanceninetypercent}[1]{\IfEqCase{#1}{{GW250114_NRSUR}{460}{GW250114_SEOB}{444}{GW250114_XO4a}{441}{GW250114_Phenom}{460}}}
\newcommand{\luminositydistancetenpercent}[1]{\IfEqCase{#1}{{GW250114_NRSUR}{347}{GW250114_SEOB}{332}{GW250114_XO4a}{324}{GW250114_Phenom}{343}}}
\newcommand{\massonesourceminus}[1]{\IfEqCase{#1}{{GW250114_NRSUR}{0.8}{GW250114_SEOB}{0.8}{GW250114_XO4a}{0.9}{GW250114_Phenom}{0.9}}}
\newcommand{\massonesourcemed}[1]{\IfEqCase{#1}{{GW250114_NRSUR}{33.6}{GW250114_SEOB}{33.5}{GW250114_XO4a}{33.5}{GW250114_Phenom}{33.7}}}
\newcommand{\massonesourceplus}[1]{\IfEqCase{#1}{{GW250114_NRSUR}{1.2}{GW250114_SEOB}{1.2}{GW250114_XO4a}{1.3}{GW250114_Phenom}{1.2}}}
\newcommand{\massonesourceninetypercent}[1]{\IfEqCase{#1}{{GW250114_NRSUR}{34.5}{GW250114_SEOB}{34.4}{GW250114_XO4a}{34.4}{GW250114_Phenom}{34.6}}}
\newcommand{\massonesourcetenpercent}[1]{\IfEqCase{#1}{{GW250114_NRSUR}{32.9}{GW250114_SEOB}{32.9}{GW250114_XO4a}{32.7}{GW250114_Phenom}{33.0}}}
\newcommand{\masstwosourceminus}[1]{\IfEqCase{#1}{{GW250114_NRSUR}{1.3}{GW250114_SEOB}{1.3}{GW250114_XO4a}{1.5}{GW250114_Phenom}{1.4}}}
\newcommand{\masstwosourcemed}[1]{\IfEqCase{#1}{{GW250114_NRSUR}{32.2}{GW250114_SEOB}{32.2}{GW250114_XO4a}{32.1}{GW250114_Phenom}{32.3}}}
\newcommand{\masstwosourceplus}[1]{\IfEqCase{#1}{{GW250114_NRSUR}{0.8}{GW250114_SEOB}{0.9}{GW250114_XO4a}{0.9}{GW250114_Phenom}{0.9}}}
\newcommand{\masstwosourceninetypercent}[1]{\IfEqCase{#1}{{GW250114_NRSUR}{32.9}{GW250114_SEOB}{32.9}{GW250114_XO4a}{32.9}{GW250114_Phenom}{33.0}}}
\newcommand{\masstwosourcetenpercent}[1]{\IfEqCase{#1}{{GW250114_NRSUR}{31.3}{GW250114_SEOB}{31.2}{GW250114_XO4a}{31.0}{GW250114_Phenom}{31.3}}}
\newcommand{\massratiominus}[1]{\IfEqCase{#1}{{GW250114_NRSUR}{0.05}{GW250114_SEOB}{0.05}{GW250114_XO4a}{0.05}{GW250114_Phenom}{0.05}}}
\newcommand{\massratiomed}[1]{\IfEqCase{#1}{{GW250114_NRSUR}{0.96}{GW250114_SEOB}{0.96}{GW250114_XO4a}{0.96}{GW250114_Phenom}{0.96}}}
\newcommand{\massratioplus}[1]{\IfEqCase{#1}{{GW250114_NRSUR}{0.04}{GW250114_SEOB}{0.04}{GW250114_XO4a}{0.04}{GW250114_Phenom}{0.04}}}
\newcommand{\massrationinetypercent}[1]{\IfEqCase{#1}{{GW250114_NRSUR}{0.99}{GW250114_SEOB}{0.99}{GW250114_XO4a}{0.99}{GW250114_Phenom}{0.99}}}
\newcommand{\massratiotenpercent}[1]{\IfEqCase{#1}{{GW250114_NRSUR}{0.91}{GW250114_SEOB}{0.91}{GW250114_XO4a}{0.91}{GW250114_Phenom}{0.91}}}
\newcommand{\redshiftminus}[1]{\IfEqCase{#1}{{GW250114_NRSUR}{0.01}{GW250114_SEOB}{0.01}{GW250114_XO4a}{0.01}{GW250114_Phenom}{0.01}}}
\newcommand{\redshiftmed}[1]{\IfEqCase{#1}{{GW250114_NRSUR}{0.09}{GW250114_SEOB}{0.08}{GW250114_XO4a}{0.08}{GW250114_Phenom}{0.09}}}
\newcommand{\redshiftplus}[1]{\IfEqCase{#1}{{GW250114_NRSUR}{0.01}{GW250114_SEOB}{0.01}{GW250114_XO4a}{0.02}{GW250114_Phenom}{0.02}}}
\newcommand{\redshiftninetypercent}[1]{\IfEqCase{#1}{{GW250114_NRSUR}{0.10}{GW250114_SEOB}{0.09}{GW250114_XO4a}{0.09}{GW250114_Phenom}{0.10}}}
\newcommand{\redshifttenpercent}[1]{\IfEqCase{#1}{{GW250114_NRSUR}{0.07}{GW250114_SEOB}{0.07}{GW250114_XO4a}{0.07}{GW250114_Phenom}{0.07}}}
\newcommand{\thetajnminus}[1]{\IfEqCase{#1}{{GW250114_NRSUR}{0.23}{GW250114_SEOB}{0.22}{GW250114_XO4a}{0.23}{GW250114_Phenom}{0.23}}}
\newcommand{\thetajnmed}[1]{\IfEqCase{#1}{{GW250114_NRSUR}{0.78}{GW250114_SEOB}{0.82}{GW250114_XO4a}{0.82}{GW250114_Phenom}{0.78}}}
\newcommand{\thetajnplus}[1]{\IfEqCase{#1}{{GW250114_NRSUR}{0.19}{GW250114_SEOB}{0.19}{GW250114_XO4a}{0.21}{GW250114_Phenom}{0.20}}}
\newcommand{\thetajnninetypercent}[1]{\IfEqCase{#1}{{GW250114_NRSUR}{0.93}{GW250114_SEOB}{0.97}{GW250114_XO4a}{0.98}{GW250114_Phenom}{0.93}}}
\newcommand{\thetajntenpercent}[1]{\IfEqCase{#1}{{GW250114_NRSUR}{0.61}{GW250114_SEOB}{0.65}{GW250114_XO4a}{0.65}{GW250114_Phenom}{0.60}}}
\newcommand{\totalmassminus}[1]{\IfEqCase{#1}{{GW250114_NRSUR}{1.0}{GW250114_SEOB}{1.1}{GW250114_XO4a}{1.2}{GW250114_Phenom}{1.1}}}
\newcommand{\totalmassmed}[1]{\IfEqCase{#1}{{GW250114_NRSUR}{71.5}{GW250114_SEOB}{71.1}{GW250114_XO4a}{70.9}{GW250114_Phenom}{71.6}}}
\newcommand{\totalmassplus}[1]{\IfEqCase{#1}{{GW250114_NRSUR}{0.9}{GW250114_SEOB}{1.1}{GW250114_XO4a}{1.3}{GW250114_Phenom}{1.0}}}
\newcommand{\totalmassninetypercent}[1]{\IfEqCase{#1}{{GW250114_NRSUR}{72.2}{GW250114_SEOB}{72.0}{GW250114_XO4a}{71.9}{GW250114_Phenom}{72.4}}}
\newcommand{\totalmasstenpercent}[1]{\IfEqCase{#1}{{GW250114_NRSUR}{70.7}{GW250114_SEOB}{70.2}{GW250114_XO4a}{69.9}{GW250114_Phenom}{70.8}}}
\newcommand{\totalmasssourceminus}[1]{\IfEqCase{#1}{{GW250114_NRSUR}{1.2}{GW250114_SEOB}{1.1}{GW250114_XO4a}{1.3}{GW250114_Phenom}{1.1}}}
\newcommand{\totalmasssourcemed}[1]{\IfEqCase{#1}{{GW250114_NRSUR}{65.8}{GW250114_SEOB}{65.7}{GW250114_XO4a}{65.5}{GW250114_Phenom}{66.0}}}
\newcommand{\totalmasssourceplus}[1]{\IfEqCase{#1}{{GW250114_NRSUR}{1.1}{GW250114_SEOB}{1.1}{GW250114_XO4a}{1.3}{GW250114_Phenom}{1.2}}}
\newcommand{\totalmasssourceninetypercent}[1]{\IfEqCase{#1}{{GW250114_NRSUR}{66.6}{GW250114_SEOB}{66.6}{GW250114_XO4a}{66.5}{GW250114_Phenom}{66.9}}}
\newcommand{\totalmasssourcetenpercent}[1]{\IfEqCase{#1}{{GW250114_NRSUR}{64.9}{GW250114_SEOB}{64.8}{GW250114_XO4a}{64.5}{GW250114_Phenom}{65.1}}}
\newcommand{\snrtwoonemultipoleminus}[1]{\IfEqCase{#1}{{GW250114_NRSUR}{0.3}{GW250114_SEOB}{0.4}{GW250114_XO4a}{0.5}{GW250114_Phenom}{0.3}}}
\newcommand{\snrtwoonemultipolemed}[1]{\IfEqCase{#1}{{GW250114_NRSUR}{0.4}{GW250114_SEOB}{0.5}{GW250114_XO4a}{0.5}{GW250114_Phenom}{0.4}}}
\newcommand{\snrtwoonemultipoleplus}[1]{\IfEqCase{#1}{{GW250114_NRSUR}{0.8}{GW250114_SEOB}{1.2}{GW250114_XO4a}{1.2}{GW250114_Phenom}{0.9}}}
\newcommand{\snrtwoonemultipoleninetypercent}[1]{\IfEqCase{#1}{{GW250114_NRSUR}{0.9}{GW250114_SEOB}{1.3}{GW250114_XO4a}{1.3}{GW250114_Phenom}{1.0}}}
\newcommand{\snrtwoonemultipoletenpercent}[1]{\IfEqCase{#1}{{GW250114_NRSUR}{0.1}{GW250114_SEOB}{0.1}{GW250114_XO4a}{0.1}{GW250114_Phenom}{0.1}}}
\newcommand{\snrthreethreemultipoleminus}[1]{\IfEqCase{#1}{{GW250114_NRSUR}{0.6}{GW250114_SEOB}{0.6}{GW250114_XO4a}{0.6}{GW250114_Phenom}{0.6}}}
\newcommand{\snrthreethreemultipolemed}[1]{\IfEqCase{#1}{{GW250114_NRSUR}{0.7}{GW250114_SEOB}{0.7}{GW250114_XO4a}{0.7}{GW250114_Phenom}{0.7}}}
\newcommand{\snrthreethreemultipoleplus}[1]{\IfEqCase{#1}{{GW250114_NRSUR}{1.2}{GW250114_SEOB}{1.2}{GW250114_XO4a}{1.3}{GW250114_Phenom}{1.2}}}
\newcommand{\snrthreethreemultipoleninetypercent}[1]{\IfEqCase{#1}{{GW250114_NRSUR}{1.6}{GW250114_SEOB}{1.7}{GW250114_XO4a}{1.7}{GW250114_Phenom}{1.6}}}
\newcommand{\snrthreethreemultipoletenpercent}[1]{\IfEqCase{#1}{{GW250114_NRSUR}{0.1}{GW250114_SEOB}{0.2}{GW250114_XO4a}{0.2}{GW250114_Phenom}{0.2}}}
\newcommand{\snrfourfourmultipoleminus}[1]{\IfEqCase{#1}{{GW250114_NRSUR}{1.5}{GW250114_SEOB}{1.5}{GW250114_XO4a}{1.6}{GW250114_Phenom}{1.6}}}
\newcommand{\snrfourfourmultipolemed}[1]{\IfEqCase{#1}{{GW250114_NRSUR}{3.6}{GW250114_SEOB}{3.9}{GW250114_XO4a}{4.0}{GW250114_Phenom}{3.6}}}
\newcommand{\snrfourfourmultipoleplus}[1]{\IfEqCase{#1}{{GW250114_NRSUR}{1.4}{GW250114_SEOB}{1.4}{GW250114_XO4a}{1.5}{GW250114_Phenom}{1.5}}}
\newcommand{\snrfourfourmultipoleninetypercent}[1]{\IfEqCase{#1}{{GW250114_NRSUR}{4.7}{GW250114_SEOB}{5.0}{GW250114_XO4a}{5.1}{GW250114_Phenom}{4.8}}}
\newcommand{\snrfourfourmultipoletenpercent}[1]{\IfEqCase{#1}{{GW250114_NRSUR}{2.4}{GW250114_SEOB}{2.7}{GW250114_XO4a}{2.7}{GW250114_Phenom}{2.4}}}
\newcommand{\eventname}{{{GW250114}}\xspace}
\newcommand{\eventnamefull}{{{GW250114\_082203}}\xspace}

\newcommand{\PaperTitle}{\eventname: Testing Hawking's Area Law and the Kerr Nature of Black Holes}

\newcommand{\Msun}{\ensuremath{M_{\odot}}\xspace}
\newcommand{\mf}{\ensuremath{M_{\rm f}}\xspace}  %
\newcommand{\chif}{\ensuremath{\chi_{\rm f}}\xspace}  %
\newcommand{\tMf}{\ensuremath{t_{M_{\rm f}}}\xspace}  %
\newcommand{\tMtot}{\ensuremath{t_{M}}\xspace}  %
\newcommand{\df}{\ensuremath{\delta f_{221}}\xspace}  %
\newcommand{\dg}{\ensuremath{\delta \gamma_{221}}\xspace}  %
\newcommand{\tRefSymbol}{\ensuremath{t_{\rm peak}}\xspace}  %
\newcommand{\tStart}{\ensuremath{t_>}\xspace}  %
\newcommand{\tEnd}{\ensuremath{t_<}\xspace}  %

\newcommand{\soft}[1]{\textsc{#1}}

\newcommand{\ringdown}{{\textsc{ringdown}}\xspace}
\newcommand{\pyring}{{\textsc{pyRing}}\xspace}
\newcommand{\cpnest}{\soft{cpnest}\xspace}
\newcommand{\Asimov}{\soft{Asimov}\xspace}
\newcommand{\BayesWave}{\soft{BayesWave}\xspace}
\newcommand{\bilby}{\soft{bilby}\xspace}
\newcommand{\rift}{\soft{RIFT}\xspace}
\newcommand{\dynesty}{\soft{dynesty}\xspace}
\newcommand{\TDInf}{\soft{TDInf}\xspace}
\newcommand{\emcee}{\soft{emcee}\xspace}
\newcommand{\arviz}{\soft{arviz}\xspace}
\newcommand{\astropy}{\soft{astropy}\xspace}
\newcommand{\cmasher}{\soft{cmasher}\xspace}

\newcommand{\jupyter}{\soft{jupyter}\xspace}
\newcommand{\ipython}{\soft{ipython}\xspace}
\newcommand{\matplotlib}{\soft{matplotlib}\xspace}

\newcommand{\pesummary}{\soft{pesummary}\xspace}

\newcommand{\qnm}{\soft{qnm}\xspace}

\newcommand{\hfivepy}{\soft{h5py}\xspace}
\newcommand{\lalsuite}{\soft{lalsuite}\xspace}
\newcommand{\swiglal}{\soft{swiglal}\xspace}
\newcommand{\gwpy}{\soft{gwpy}\xspace}
\newcommand{\pandas}{\soft{pandas}\xspace}
\newcommand{\scipy}{\soft{scipy}\xspace} 
\newcommand{\seaborn}{\soft{seaborn}\xspace}  
\newcommand{\tqdm}{\soft{tqdm}\xspace} 
\newcommand{\numpy}{\soft{numpy}\xspace}
\newcommand{\python}{\soft{python}\xspace}
\newcommand{\pySEOBNR}{\soft{pySEOBNR}\xspace}
\newcommand{\numpyro}{\soft{numpyro}\xspace}
\newcommand{\jax}{\soft{jax}\xspace}

\newcommand{\GSTLAL}{\soft{GstLAL}\xspace}
\newcommand{\CWB}{\soft{cWB}\xspace}
\newcommand{\PYCBC}{\soft{PyCBC}\xspace}
\newcommand{\MBTA}{\soft{MBTA}\xspace}
\newcommand{\SPIIR}{\soft{SPIIR}\xspace}
\newcommand{\MLy}{\soft{MLy}\xspace}
\newcommand{\mPrimary}{{\textcolor{black}{\massonesourcemed{GW250114_NRSUR}^{+\massonesourceplus{GW250114_NRSUR}}_{-\massonesourceminus{GW250114_NRSUR}}}}}
\newcommand{\mSecondary}{{\textcolor{black}{\masstwosourcemed{GW250114_NRSUR}^{+\masstwosourceplus{GW250114_NRSUR}}_{-\masstwosourceminus{GW250114_NRSUR}}}}}
\newcommand{\Mtotaltab}{{\textcolor{black}{\totalmasssourcemed{GW250114_NRSUR}^{+\totalmasssourceplus{GW250114_NRSUR}}_{-\totalmasssourceminus{GW250114_NRSUR}}}}}
\newcommand{\massratio}{{\geq \massratiotenpercent{GW250114_NRSUR}}}
\newcommand{\sPrimary}{{\textcolor{black}{\leq \aoneninetypercent{GW250114_NRSUR}}}}
\newcommand{\sSecondary}{{\textcolor{black}{\leq \atwoninetypercent{GW250114_NRSUR}}}}
\newcommand{\Mfinal}{{\textcolor{black}{\finalmasssourcemed{GW250114_NRSUR}^{+\finalmasssourceplus{GW250114_NRSUR}}_{-\finalmasssourceminus{GW250114_NRSUR}}}}}
\newcommand{\chifinal}{\ensuremath{\textcolor{black}{\finalspinmed{GW250114_NRSUR}^{+\finalspinplus{GW250114_NRSUR}}_{-\finalspinminus{GW250114_NRSUR}}}}\xspace}
\newcommand{\mulitpoleSNR}{{\textcolor{black}{\snrfourfourmultipolemed{GW250114_NRSUR}^{+\snrfourfourmultipoleplus{GW250114_NRSUR}}_{-\snrfourfourmultipoleminus{GW250114_NRSUR}}}}}

\newcommand{\massconstrain}{{\textcolor{black}{2}}\xspace}
\newcommand{\improvement}{{\textcolor{black}{3{-}4}}\xspace}

\newcommand{\EccUpperLimit}{{\textcolor{black}{\leq \SEOBBilbyEccentricityUpperLimit}}\xspace}
\newcommand{\EccRefFreq}{{\textcolor{black}{\EccentricityRefFreq}}\xspace}

\newcommand{\JSNRSurCalibration}{{\textcolor{black}{0.002}}\xspace}

\newcommand{\JSNRSurvsPhenom}{{\textcolor{black}{0.018}}\xspace}
\newcommand{\JSNRSurvsEOB}{{\textcolor{black}{0.041}}\xspace}
\newcommand{\JSNRSurvsPhenomOfour}{{\textcolor{black}{0.097}}\xspace}

\newcommand{\NRSur}{{\textcolor{black}{\textsc{NRSur7dq4}}}\xspace}
\newcommand{\Phenom}{{\textcolor{black}{\textsc{PhenomXPHM}}}\xspace}
\newcommand{\EOB}{{\textcolor{black}{\textsc{SEOBNRv5PHM}}}\xspace}
\newcommand{\PhenomOfour}{{\textcolor{black}{\textsc{PhenomXO4a}}}\xspace}

\newcommand{\EOBEcc}{{\textcolor{black}{\textsc{SEOBNRv5EHM}}}\xspace}
\newcommand{\TEOBEcc}{{\textcolor{black}{\textsc{TEOBResumS-DALI}}}\xspace}

\newcommand{\SNR}{{\textcolor{black}{80}}\xspace}
\newcommand{\SNRFifteenZeroNineFourteen}{{\textcolor{black}{26}}\xspace}
\newcommand{\SNRTwentyThreeZeroEightNineteen}{{\textcolor{black}{42}}\xspace}

\newcommand{\duration}{{\textcolor{black}{8}}\xspace}
\newcommand{\fmin}{{\textcolor{black}{20}}\xspace}
\newcommand{\fmax}{{\textcolor{black}{896}}\xspace}
\newcommand{\FrequencyRange}{\textcolor{black}{\([\fmin, \fmax]\)\,Hz}\xspace}

\newcommand{\AmpCalib}{{\textcolor{black}{4\%}}\xspace}
\newcommand{\PhaseCalib}{{\textcolor{black}{2}}\xspace}
\newcommand{\EllThreeFrequency}{{\textcolor{black}{22.5}}\xspace}
\newcommand{\EllFourFrequency}{{\textcolor{black}{30}}\xspace}

\newcommand{\BayesWaveOverlap}{\textcolor{black}{0.995}\xspace}

\newcommand{\ALsign}{{\ensuremath{4.4\sigma}}\xspace}
\newcommand{\ALsignminimum}{{{\ensuremath{3.4\sigma}}}\xspace}
\newcommand{\Pretime}{{{-40}}}
\newcommand{\PreSNR}{{{55}}\xspace}

\newcommand{\ALsignOvertone}{{3.6}}
\newcommand{\AreaLawFiveSigmaTime}{{-10}}

\newcommand{\FluxPeak}{{{6}}}
\newcommand{\FluxOnePercent}{{{-232}}}
\newcommand{\FluxTenPercent}{{{-36}}}

\newcommand{\TDinfSeglen}{{1.4}}
\newcommand{\TDinfTimeStd}{{0.01}}
\newcommand{\Specsign}{{{\pm{30\%}}}\xspace}
\newcommand{\tFundamental}{{{27.0}}\xspace}  %
\newcommand{\tFundamentalEarly}{{{20}}\xspace}  %
\newcommand{\tFundamentalSNR}{{{10.5}}\xspace}  %
\newcommand{\SNRFundamentalEarly}{{{21}}\xspace}  %
\newcommand{\tOvertone}{{{10.5}}\xspace}  %
\newcommand{\tSecondOvertone}{{{6}}\xspace}  %

\newcommand{\OvertoneThreeSigmaTime}{\ensuremath{{9.0}}\xspace}
\newcommand{\OvertoneSignificance}{\ensuremath{{4.1\sigma}}\xspace}
\newcommand{\FundamentalSignificance}{\ensuremath{{{>}{7}\sigma}}\xspace}
\newcommand{\PyRingDelay}{{0.5}\xspace}

\newcommand{\tRef}{{{1420878141.235932}}\xspace}
\newcommand{\tRefH}{{{1420878141.2190118}}\xspace}
\newcommand{\tRefL}{{{1420878141.2165165}}\xspace}
\newcommand{\raRef}{{{2.333}}\xspace}
\newcommand{\decRef}{{{0.190}}\xspace}
\newcommand{\psiRef}{{{1.329}}\xspace}
\newcommand{\mfRef}{{{68.409}}\xspace}
\newcommand{\tMfRef}{{{0.337}\, {\rm ms}}\xspace}

\newcommand{\mRef}{\ensuremath{{71.849}\, \Msun}\xspace}
\newcommand{\tMRef}{\ensuremath{{0.354}\, {\rm ms}}\xspace}

\newcommand{\tPeakH}{{{1420878141.2190^{+0.0001}_{-0.0001}}\, {\rm s}}\xspace}
\newcommand{\tPeakL}{{{1420878141.2165^{+0.0001}_{-0.0001}}\, {\rm s}}\xspace}
\newcommand{\tPeakHrel}{{{-0.07^{+0.37}_{-0.37}}\, \tMf}\xspace}
\newcommand{\tPeakLrel}{{{0.07^{+0.38}_{-0.38}}\, \tMf}\xspace}
\newcommand{\tPeakHrelStd}{{{0.22}\, \tMf}\xspace}
\newcommand{\tPeakLrelStd}{{{0.23}\, \tMf}\xspace}
\newcommand{\tPeakOverallCI}{{{0.4}\, \tMf}\xspace}

\newcommand{\dfConstraint}{{{\ensuremath{0.1^{+0.3}_{-0.3}}}}\xspace}

\newcommand{\FundamentalFreq}{\ensuremath{{247^{+6}_{-6}}}\xspace}
\newcommand{\FundamentalDamp}{\ensuremath{{221^{+39}_{-32}}}\xspace}
\newcommand{\OvertoneFreq}{\ensuremath{{249^{+8}_{-9}}}\xspace}
\newcommand{\OvertoneDamp}{\ensuremath{{708^{+116}_{-107}}}\xspace}
\newcommand{\OvertoneSNR}{\ensuremath{{26}}\xspace}

\newcommand{\PyRingSeglen}{{64}}

\newcommand{\TimingDiscussionFundamentalTimeRingdown}{{10.5}}
\newcommand{\tGPSringdownFundamentalRefH}{{1420878141.222534}}
\newcommand{\tGPSringdownFundamentalRefL}{{1420878141.220032}}
\newcommand{\tGPSringdownFundamentalRefHdelta}{{-0.046}}
\newcommand{\tGPSringdownFundamentalRefLdelta}{{-0.067}}

\newcommand{\TimingDiscussionFundamentalTimePyring}{{10}}
\newcommand{\tGPSpyringFundamentalRefH}{{1420878141.222412}}
\newcommand{\tGPSpyringFundamentalRefL}{{1420878141.219971}}
\newcommand{\tGPSpyringFundamentalRefHdelta}{{-0.41}}
\newcommand{\tGPSpyringFundamentalRefLdelta}{{-0.25}}

\newcommand{\tGPSfundamentalPyringRingdownH}{{-0.36}}
\newcommand{\tGPSfundamentalPyringRingdownL}{{-0.18}}

\newcommand{\TimingDiscussionOvertoneTimeRingdown}{6}
\newcommand{\TimingDiscussionOvertoneTimePyring}{6}
\newcommand{\tGPSpyringOvertoneRefDelta}{{0.54}}
\newcommand{\networkMatchedFilterSNRlower}{{77}\xspace}
\newcommand{\networkMatchedFilterSNRupper}{{80}\xspace}
\newcommand{\HanfordMatchedFilterSNR}{{53}\xspace}
\newcommand{\LivingstonMatchedFilterSNR}{{60}\xspace}

\newcommand{\ApproximateStrainAmplitude}{{\ensuremath{\sim 10^{-21}}}\xspace}

\newcommand{\TimeDelay}{{75}\xspace}
\newcommand{\FARThreshold}{{1 per 100}\xspace}

\newcommand{\SkymapTime}{{08:22:35 UTC}\xspace}
\newcommand{\SkymapTimeDelay}{{32}\xspace}
\newcommand{\PBlackHoleLowerBound}{{99}\%\xspace}

\newcommand{\GSTLALTimeDelay}{{15}\xspace}

\newcommand{\utctimeDateFirst}{{January 14th, 2025, 08:22:03 UTC}\xspace}
\newcommand{\utctimeTimeFirst}{{08:22:03 UTC on January 14th, 2025}\xspace} %

\newcommand{\SEOBBilbyEccentricityUpperLimit}{\ensuremath 0.03}

\newcommand{\EccentricityRefFreq}{\ensuremath 13.33}

\def\gitid{%
P2500421-v8-72-g3b7d771f-dirty
}
\let\keeptoday\today
\def\today{\keeptoday, GitID {\normalfont\gitid}}

\usepackage{siunitx}
\DeclareSIUnit\parsec{pc}
\DeclareSIUnit\Mpc{\mega\parsec}

\usepackage{txfonts}
\newlength{\capheight}
\graphicspath{{source/figures/}}

\begin{document}

\title{\PaperTitle}

\ifprintauthors
\author{A.~G.~Abac\,\orcidlink{0000-0003-4786-2698}}
\affiliation{Max Planck Institute for Gravitational Physics (Albert Einstein Institute), D-14476 Potsdam, Germany}
\author{I.~Abouelfettouh}
\affiliation{LIGO Hanford Observatory, Richland, WA 99352, USA}
\author{F.~Acernese}
\affiliation{Dipartimento di Farmacia, Universit\`a di Salerno, I-84084 Fisciano, Salerno, Italy}
\affiliation{INFN, Sezione di Napoli, I-80126 Napoli, Italy}
\author{K.~Ackley\,\orcidlink{0000-0002-8648-0767}}
\affiliation{University of Warwick, Coventry CV4 7AL, United Kingdom}
\author{C.~Adamcewicz\,\orcidlink{0000-0001-5525-6255}}
\affiliation{OzGrav, School of Physics \& Astronomy, Monash University, Clayton 3800, Victoria, Australia}
\author{S.~Adhicary\,\orcidlink{0009-0004-2101-5428}}
\affiliation{The Pennsylvania State University, University Park, PA 16802, USA}
\author{D.~Adhikari}
\affiliation{Max Planck Institute for Gravitational Physics (Albert Einstein Institute), D-30167 Hannover, Germany}
\affiliation{Leibniz Universit\"{a}t Hannover, D-30167 Hannover, Germany}
\author{N.~Adhikari\,\orcidlink{0000-0002-4559-8427}}
\affiliation{University of Wisconsin-Milwaukee, Milwaukee, WI 53201, USA}
\author{R.~X.~Adhikari\,\orcidlink{0000-0002-5731-5076}}
\affiliation{LIGO Laboratory, California Institute of Technology, Pasadena, CA 91125, USA}
\author{V.~K.~Adkins}
\affiliation{Louisiana State University, Baton Rouge, LA 70803, USA}
% ...
 \fi
\collaboration{The LIGO Scientific Collaboration, the Virgo Collaboration, and the KAGRA Collaboration}

\email{lvc.publications@ligo.org}

\begin{abstract}

The gravitational-wave signal \eventname was observed by the two LIGO detectors with a network matched-filter signal-to-noise ratio of \SNR. 
The signal was emitted by the coalescence of two black holes with near-equal masses $m_1=\mPrimary\, \Msun$ and $m_2=\mSecondary\, \Msun$, and small spins $\chi_{1,2}\, \sSecondary$ (90\% credibility) and negligible eccentricity $e\ \EccUpperLimit$.
Post-merger data excluding the peak region are consistent with the dominant quadrupolar $(\ell=|m|=2)$ mode of a Kerr black hole and its first overtone.
We constrain the modes' frequencies to $\Specsign$ of the Kerr spectrum, providing a test of the remnant's Kerr nature.
We also examine Hawking's area law, also known as the second law of black hole mechanics, which states that the total area of the black hole event horizons cannot decrease with time.
A range of analyses that exclude up to five of the strongest merger cycles confirm that the remnant area is larger than the sum of the initial areas to high credibility.
\end{abstract}

\maketitle

\section{Introduction}

\begin{figure*}
    \centering
    \includegraphics[width=\linewidth]{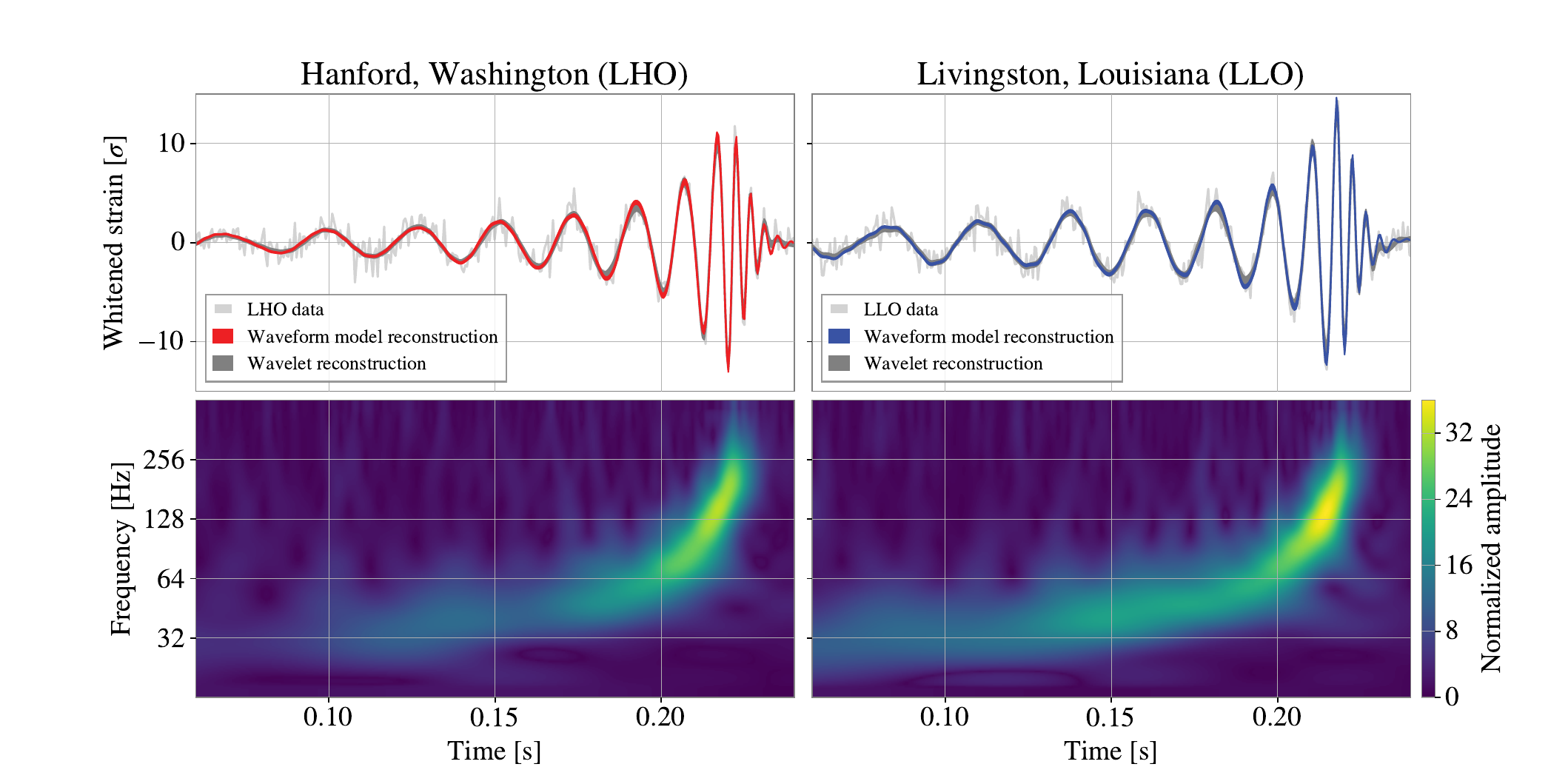}
    \caption{
    Data from LIGO Hanford (left) and LIGO Livingston (right) and \eventname signal reconstruction. 
    Times are relative to \utctimeDateFirst. 
    The top panels show whitened data versus time and signal reconstructions (90\% credible regions), either with a waveform model for black hole binaries in general relativity~\cite{Varma:2019csw} or via a model-agnostic wavelet-based approach~\cite{Cornish:2020dwh, Littenberg:2014oda}.
    Data and models have been downsampled to $2048$\,Hz, whitened (effectively, divided) by the detector noise amplitude spectral density, and finally bandpassed to \FrequencyRange.
    The bottom panels show a time\textendash frequency spectrogram of the data.
    The signal reaches ${>}10\sigma$ above the noise.
    }
    \label{fig:reconstruction}
\end{figure*}

Ten years after the detection of GW150914~\cite{LIGOScientific:2016aoc} by the LIGO detectors~\cite{LIGOScientific:2014pky}, gravitational-wave astronomy is thriving.
Advances in detector performance~\cite{Capote:2024rmo}, including breakthroughs in quantum precision measurement \cite{VIRGO:2023wes,LIGOO4Detector:2023wmz,membersoftheLIGOScientific:2024elc},
have led the field from the first glimpse of merging black holes~\cite{LIGOScientific:2016lio,LIGOScientific:2016vpg} to an expanding catalog of hundreds of detections~\citep{GWTC1:2018mvr, Zackay:2019btq, GWTC2.1:2021usb, Nitz:2020oeq,  Nitz:2021uxj, GWTC3:2021vkt, Olsen:2022pin, Nitz:2021zwj, Mehta:2023zlk, GWTC4results}.
Here we report the observation of \eventnamefull, henceforth \eventname, shown in Fig.~\ref{fig:reconstruction}.
With similar parameters to GW150914, \eventname reaches a similar strain amplitude of \ApproximateStrainAmplitude. 
Yet, thanks to the LIGO detectors now operating near their design sensitivity~\cite{Capote:2024rmo}, it registers at a signal-to-noise ratio of \SNR, as opposed to \SNRFifteenZeroNineFourteen for GW150914 a decade ago.
This makes \eventname the most clearly recorded signal to date,
broadening the scope of fundamental tests of strong-field gravity and black holes.

From a theoretical standpoint, black holes are expected to be remarkably simple objects~\cite{Israel:1967wq,Israel:1967za,Carter:1971zc,Hawking:1971vc,Robinson:1975bv,Mazur:1982db,Bunting:1983,Chrusciel:2023onh,Alexakis:2010nsf}.
According to Einstein's theory of general relativity and 
under suitable regularity assumptions, isolated stationary black holes can be fully characterized by just three parameters: mass, spin, and electromagnetic charge.
For neutral black holes, this implies that mass and spin determine the system through the Kerr metric \cite{Kerr:1963ud}, the unique axisymmetric, neutral solution to Einstein's equations \cite{Robinson:1975bv}.
This uniqueness is closely tied to key conjectures in classical gravitation, including weak cosmic censorship \cite{Penrose:1969pc} and the stability of rotating black holes \cite{Vishveshwara:1970cc,Kay:1987ax,Whiting:1988vc,Klainerman:2021qzy,Dafermos:2016uzj,TeixeiradaCosta:2019skg,Dafermos:2021cbw}, both of which remain unproven.

The uniqueness and implied featurelessness of black holes gives rise to paradoxes in the context of quantum mechanics and thermodynamics~\cite{Wald:1999vt}.
The laws of black hole mechanics,
originally suspected to be only coincidentally reminiscent of statistical mechanics~\cite{Bardeen:1973gs}, establish black holes as true thermodynamic systems~\cite{Wald:1999vt,PhysRevD.48.R3427,Jacobson:1993vj}:
the role of the entropy is assigned to the area of the event horizon~\cite{PhysRevD.7.2333,PhysRevD.9.3292}, while black holes radiate due to quantum effects as a black body with a temperature related to their surface gravity~\cite{hawking1974blackhole}.
Black hole thermodynamics plays a key role in the quest to reconcile gravity with the rest of physics~\cite{Bousso:2022ntt}, through concepts such as information loss~\cite{Almheiri:2012rt}, holographic gravity~\cite{tHooft:1993dmi,Maldacena:1997re}, or the microscopic interpretation of black hole entropy~\cite{PhysRevD.7.2333,Strominger:1996sh, Sen:1996pb}.

Black holes are not just mathematical idealizations.
They play a central role in the phenomenology and evolution of the Universe, displaying rich and complex behaviors from stellar to galactic scales~\cite{Cattaneo:2009ub,EventHorizonTelescope:2019dse,EventHorizonTelescope:2022wkp,Ghez:2008ms,Gillessen:2008qv,Motta:2021zlt,Remillard:2006fc,Webster:1972bsw,1972Natur.235..271B,Schmidt:1963wkp,Gaia:2024ggk}.
Astrophysical black holes are expected to not be significantly charged \cite{Gibbons:1975kk,Wald:1974np,Zajacek:2019kla}, thus conforming to the Kerr metric---yet, the extent to which they do so is an open question.
Gravitational waves can inform this by observationally probing the Kerr nature of black holes.

Although black hole coalescences feature some of the strongest and most dynamical gravitational fields, their initial and final states are simple.
A coalescence begins with a long inspiral, during which two black holes orbit and approach each other as the system loses energy and angular momentum to gravitational radiation \cite{PhysRev.131.435,Blanchet:2013haa,LIGOScientific:2016wyt}.
After the merger, the remnant black hole ``rings"~\cite{Vishveshwara:1970zz} as it settles into a quiescent Kerr state \cite{Doroshkevich:1966,Price:1971fb,Price:1972pw}.
In this context, assuming a Kerr remnant implies a specific ringdown spectrum (frequencies and damping rates) which is a known function of the black hole mass and spin~\cite{Berti:2009kk}. 
In parallel, the second law of black hole mechanics, also known as Hawking's area law, requires a net increase in the total event horizon area throughout the coalescence~\cite{PhysRevLett.26.1344}.

\eventname enables precise tests of both Hawking's area law and the Kerr nature of black holes.
Excluding the neighborhood of the signal peak, we establish that the post-merger data contain at least two distinct ringing modes of the remnant at the \OvertoneSignificance credible level.
These modes are consistent with the fundamental and first overtone of the quadrupolar $(\ell=|m|=2)$ spectrum of a Kerr black hole; deviations in the mode frequencies are constrained to $\Specsign$.
To test the area law, we infer the initial black hole areas using pre-merger data that exclude up to five merger signal cycles and the final black hole area using post-merger data with one-mode or two-mode models at their earliest time of applicability.
In all cases, the remnant's event horizon area exceeds the total initial area at high credibility, in agreement with Hawking’s law.
Thanks to the strength of \eventname, this test is possible even when excising the loudest portion of the signal where gravity is at its strongest and most dynamical.

\section{Observation of GW250114}

\begin{figure}
    \centering
    \includegraphics[width=\linewidth]{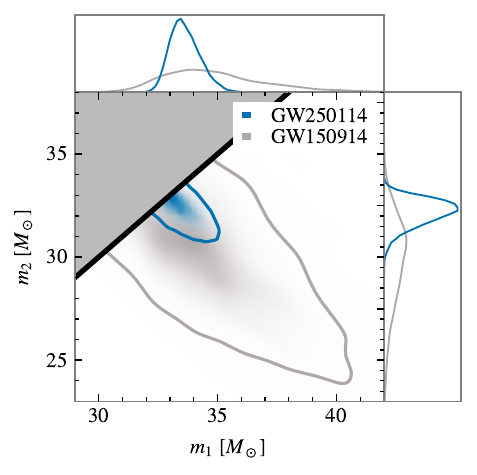}
    \caption{Posterior distribution for the source-frame component masses of \eventname, under the definition $m_1 \geq m_2$ (marginals and 90\%-credible contours).
    For comparison we also show results for GW150914~\cite{LIGOScientific:2016vlm,GWTC2.1:2021usb}.}
    \label{fig:PE_mass}
\end{figure}

\eventname arrived at \utctimeTimeFirst, while the LIGO Hanford and LIGO Livingston detectors~\cite{LIGOScientific:2014pky} were operating nominally, Virgo~\cite{VIRGO:2014yos} was undergoing routine maintenance, and KAGRA~\cite{KAGRA:2020tym} was not taking data.
No significant data-quality issues were identified at the time~\cite{LIGO:2024kkz}.
\eventname was detected with high significance by all search pipelines operating at the time; \GSTLAL~\cite{Cannon:2011rj, Messick:2016aqy, Sachdev:2019vvd, Hanna:2019ezx, Cannon:2020qnf, Tsukada:2023edh, Ewing:2023qqe, Sakon:2022ibh, Joshi:2025nty, Joshi:2025zdu, Joshi:2023ltf, Ray:2023nhx}, \MLy~\cite{Skliris:2020qax}, \SPIIR~\cite{Chu:2020pjv}, \MBTA~\cite{Aubin:2020goo,Allene:2025saz}, \PYCBC~\cite{DalCanton:2020vpm}, and \CWB~\cite{Klimenko:2015ypf,Mishra:2024zzs}, with a network matched-filter signal-to-noise ratio, henceforth SNR,~\cite{PhysRevD.44.3819,Allen:2005fk} ranging between \networkMatchedFilterSNRlower and \networkMatchedFilterSNRupper. 
This is the highest reported SNR to date, surpassing GW230814\_230901 which had SNR \SNRTwentyThreeZeroEightNineteen~\cite{GW230814,GWTC4results}.
\eventname's record extends to the individual detectors, with SNRs of \HanfordMatchedFilterSNR and \LivingstonMatchedFilterSNR at LIGO Hanford and LIGO Livingston respectively.
The combination of the two LIGO detectors at comparable sensitivity enables the measurement precision reported here.

We infer the source properties following standard procedures~\cite{GWTC4methods}. 
Here we quote selected results obtained with the \NRSur waveform model~\citep{Varma:2019csw}, a surrogate of vacuum numerical-relativity simulations of spin-precessing, quasicircular systems. We obtain consistent results with other models~\citep{Pratten:2020ceb, Colleoni:2024knd, Pompili:2023tna, Khalil:2023kep, vandeMeent:2023ols, Ramos-Buades:2023ehm, Hamilton:2021pkf, Thompson:2023ase, Gamboa:2024hli, Nagar:2024dzj}, presented with further technical details in the Supplement\citesupp.
The source of \eventname has a binary total mass $M = \Mtotaltab\,\Msun$ and mass ratio $q\, \massratio$.
Above and throughout, all quantities are quoted at the 90\% credible level unless stated otherwise.
We constrain the component masses to within ${\sim}\massconstrain \, M_\odot$, a factor of $\improvement$ improvement compared to GW150914~\cite{LIGOScientific:2016aoc}, at $m_1=\mPrimary\, \Msun,\, m_2=\mSecondary\, \Msun$; see Fig.~\ref{fig:PE_mass}. 
The black hole dimensionless spins, $\chi\equiv Sc/(G m^2)$ where $S$ is the spin angular momentum, are both small, $\chi_{1}\, \sPrimary$,  $\chi_{2} \, \sSecondary$, with no evidence for precession. 
The remnant black hole has a mass of $\mf=\Mfinal\,M_{\odot}$ and a spin of $\chif=\chifinal$.
Furthermore, there is support for the $\ell=|m|=4$ radiation multipole with a network SNR of $ \mulitpoleSNR$. 
The source parameters are consistent with GW150914 and the wider population of binary black hole mergers~\citep{GWTC3:2021vkt,GWTC4pop}, which contains an overdensity of observed black holes in the $30{-}40\, M_\odot$ mass range and small spins~\citep{GWTC3pop:2021duu, GWTC4pop}.
The model-based signal reconstruction is consistent with a model-agnostic, wavelet-based approach~\cite{Cornish:2020dwh}, to within their statistical uncertainties, see Fig.~\ref{fig:reconstruction}.
Their noise-weighted overlap~\cite{LIGOScientific:2019hgc,Ghonge:2020suv} is \BayesWaveOverlap.

A complementary analysis with models that allow for eccentric orbits but non-precessing spins~\cite{Gamboa:2024hli,Nagar:2024dzj} finds that the eccentricity is constrained to $e\ \EccUpperLimit$ at a gravitational-wave frequency of $\EccRefFreq\,$Hz.

\section{The ringing of the remnant black hole}

\emph{Background.} The merger process gives rise to a distorted black hole that rings down into quiescence \cite{Kerr:1963ud,Doroshkevich:1966,Price:1971fb,Price:1972pw,Vishveshwara:1970zz,Campanelli:2008dv}.
Modeling this ringdown signal is a key ingredient of probes of the Kerr nature of the remnant and the area law.
Within the framework of perturbation theory, the remnant signal is dominated by a superposition of quasinormal modes of the form  $h \sim \exp(-2\pi i f t - \gamma t)$ \cite{Vishveshwara:1970cc,Press:1971wr,Davis:1972ud,Teukolsky:1973ha,Chandrasekhar:1975zza}.
Each mode's frequency $f$ and damping rate $\gamma$ are determined by the asymptotic Kerr mass \mf and spin \chif, while its amplitude and phase depend on the details of the coalescence.
Since the only relevant scale is the black hole mass, heavier black holes ring at lower frequencies and for a longer time; for a given mass, spin generally increases the damping time.
While the spectrum depends on the assumptions of general relativity and a Kerr metric (with ingoing and outgoing boundary conditions at the black hole and infinity, respectively), quasinormal modes arise generically also in alternative frameworks~\cite{Blazquez-Salcedo:2017txk,Hussain:2022ins,Chung:2024ira,Chung:2025gyg,Cardoso:2018ptl,Cano:2023jbk,Corman:2022xqg,Cayuso:2023xbc,Pierini:2021jxd,Pierini:2022eim,Wagle:2021tam,Li:2025fci}.
Detected frequencies and times are scaled by a cosmological redshift factor, and the corresponding redshifted timescale is $\tMf = \left(1+z\right) G \mf /c^3$.

Individual modes are indexed by angular numbers $(\ell, m)$ and an overtone number $n$.
For a fixed $(\ell, m)$, modes with a higher $n$ typically have higher damping rates \cite{Cook:2014cta}.
The quadrupolar geometry of the binary and the fact that gravitational radiation is leading-order quadrupolar \cite{1982ApJ...253..908T, Kramer:2021jcw,LIGOScientific:2021sio} mean that in equal-mass, nonprecessing, quasicircular binaries similar to \eventname, prograde modes with $\ell = |m| = 2$ dominate the signal.
The longest-lived, fundamental, $n=0$ mode is the main contributor at late times.
Fits to numerical-relativity simulations of similar systems additionally suggest that the next strongest mode is $n=1$, and that it decays below the fundamental around $10\,\tMf$ after the merger~\cite{Buonanno:2006ui,London:2014cma,Giesler:2019uxc,Cheung:2023vki,Giesler:2024hcr,Pacilio:2024tdl,MaganaZertuche:2024ajz,Mitman:2025hgy}.

\emph{Method.} We model the post-merger signal from \eventname using black hole perturbation theory, representing it as a superposition of generically polarized damped sinusoids~\cite{Isi:2021iql,Isi:2022mbx}.
This analysis is distinct from the full inspiral\textendash merger\textendash ringdown treatment described earlier, which incorporates numerical relativity to capture the complex merger dynamics.
Our objective here is to directly test the predictions of first-order perturbation theory against the data.

We choose a reference timescale and merger time consistent with the full-signal analysis.
We adopt a reference timescale of $\tMf = \tMfRef$.
The reference time $\tRefSymbol$ is defined via the peak of the inferred strain amplitude \cite{Varma:2019csw} and is measured to within $\pm\tPeakOverallCI$ at each detector.
Both \tMf and \tRefSymbol are only reference points and not a true dependence of the analysis.
Using the time-domain \ringdown \cite{Isi:2021iql} and \pyring \cite{Carullo:2019flw} inference packages, we model the data after some start time $\tStart$ with different mode combinations and obtain posteriors for their parameters.
The start time \tStart is reported relative to \tRefSymbol.
See the Supplement for details\citesupp.

\begin{figure*}
\centering
\includegraphics[width=\linewidth]{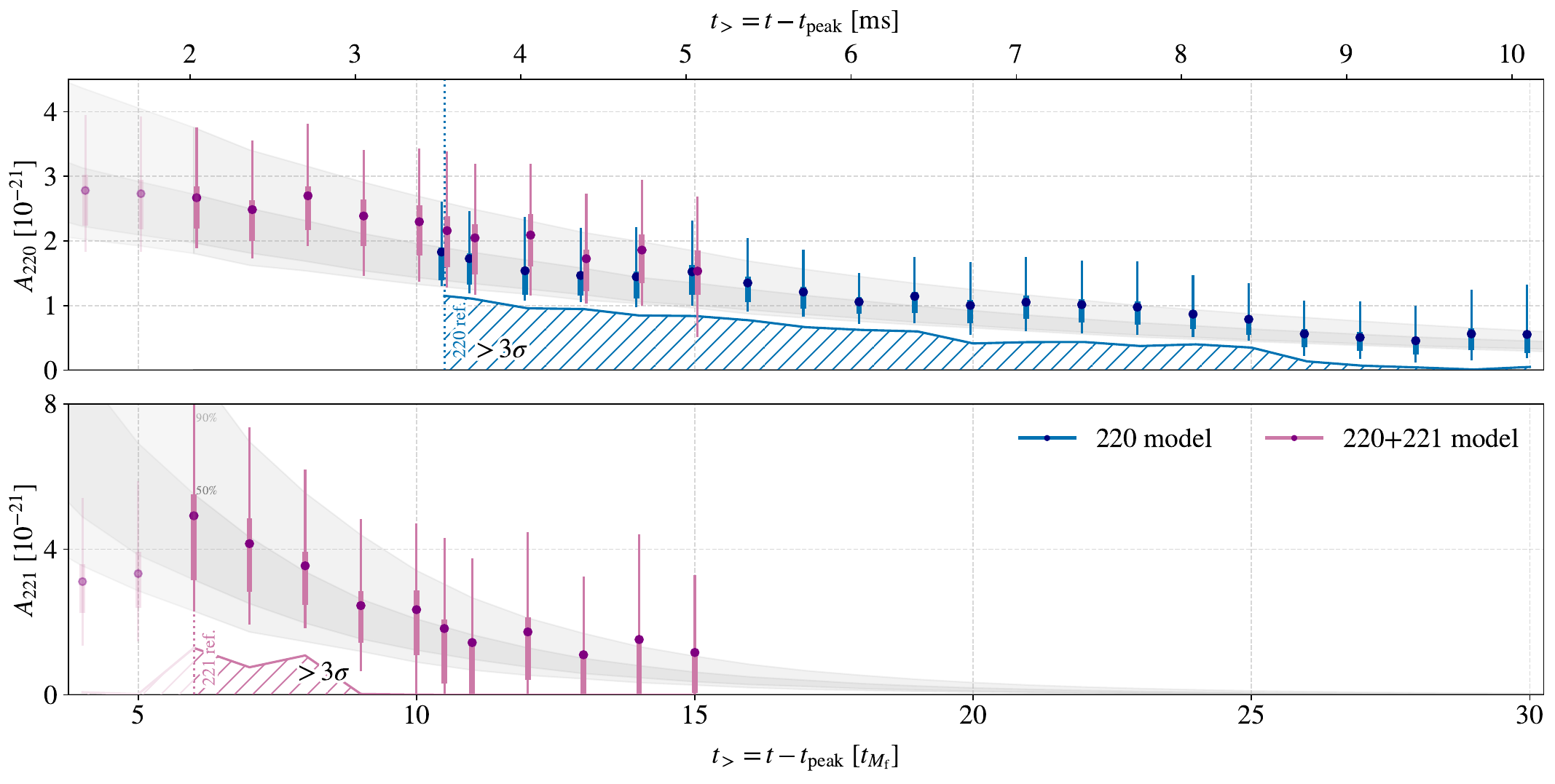}
\caption{
Strain amplitudes of the fundamental mode $A_{220}$ (top) and first overtone $A_{221}$ (bottom) reported at the start of each analysis, $\tStart=t-\tRefSymbol$, 
in units of $\tMf$ (bottom $x$-axis) and milliseconds (top $x$-axis). 
Bars mark 90\% (thin) and 50\% (thick) credible intervals around the median (circle), when modeling the data with one (blue) or two (pink) modes. Gray shading shows the expected amplitude decay as inferred from $\tStart = \tOvertone\, \tMf$ for the single-mode analysis (top; 220 ref), and $\tStart = \tSecondOvertone\, \tMf$ for the two-mode analysis (bottom; 221 ref).
Single-mode results are shown only after $\tOvertone\, \tMf$ (top), where support for a second mode is $<1\sigma$.
Overtone results are shown up to $15\, \tMf$, by which time support for a second mode has completely vanished.
Results pre-$\tSecondOvertone\, \tMf$ are given lower opacity as the overtone decay does not follow the expected trend.
Hatched regions are excluded to at least $3\sigma$; higher significance regions are contained within it.
We plot \ringdown results but obtain qualitatively consistent results with \pyring.
Mode frequencies, damping rates and SNRs are plotted in the Supplement\citesupp.
}
\label{fig:QNMamplitudes}
\end{figure*}

\emph{Modes in \eventname.} Since perturbation theory refers to the asymptotic spacetime, we start by analyzing the data from late times and seek the earliest time after which the signal can be described by a single mode;
see Fig.~\ref{fig:QNMamplitudes}, blue.
The mode frequency and damping rate are parametrized in terms of a Kerr black hole mass and spin; this is equivalent to directly parametrizing in terms of frequency and damping rate as in Ref.~\cite{LIGOScientific:2016lio} up to the choice of priors~\cite{Isi:2021iql}.
We find that data at late times are consistent with a single damped sinusoid, whose amplitude is confidently constrained away from zero at \FundamentalSignificance credibility for start times as late as $\tStart = \tFundamentalEarly\, \tMf$.
The mode decays away, falling below $3\sigma$ credibility at $\tStart = \tFundamental\, \tMf$.
The inferred mode amplitude evolves consistently with a damped sinusoid in noise (gray shading).
As discussed above, symmetry and late-time arguments suggest that this mode is the fundamental $(\ell=|m|=2,n=0)$ mode of the remnant black hole, without any quantitative comparison to the full-signal analysis \cite[e.g.,][]{LIGOScientific:2016lio}---we label the mode as such in the top panel of Fig.~\ref{fig:QNMamplitudes}.
Starting at $\tStart = \tFundamentalSNR\, \tMf$, this model recovers a SNR of $\SNRFundamentalEarly$ (Fig.~7 in Supplement\citesupp) and infers the mode's (redshifted) frequency and damping rate to be $f_{220} = \FundamentalFreq\, {\rm Hz}$ and $\gamma_{220} = \FundamentalDamp\, {\rm Hz}$, respectively.
 
Pushing $\tStart$ earlier in time and toward the merger increases the SNR, but risks contamination from other modes, linear effects beyond exponentially damped sinusoids, or nonlinear effects~\cite{London:2014cma,Sberna:2021eui,Mitman:2022qdl,Cheung:2022rbm,Chavda:2024awq,DeAmicis:2025xuh}.
The nonorthogonality of damped sinusoids further means that a model with additional modes does not necessarily lead to more faithful inference.
We therefore again start from late times and seek (i) the time a second mode is required and (ii) the earliest time after which the data agree with two modes whose amplitudes decay self-consistently.
Given the identification of a single damped sinusoid at late times, we expect any additional modes to be short-lived.
Following Refs.~\cite{Isi:2019aib,CalderonBustillo:2020rmh,Cotesta:2022pci,Isi:2022mhy,Cotesta:2022pci,Isi:2023nif,Finch:2022ynt,Correia:2023bfn,Wang:2023xsy,Ma:2023vvr,LIGOScientific:2020tif,LIGOScientific:2021sio} and expectations for \eventname-like systems~\cite{Giesler:2024hcr,Mitman:2025hgy, Gao:2025zvl}, we enhance the model with the first overtone ($\ell = |m| =2, n=1$), and parametrize the frequencies and damping rates of both modes as a function of a single mass and spin.

Figure \ref{fig:QNMamplitudes} shows two-mode results in pink.
In the bottom panel, we confidently extract the overtone for a range of start times.
Its amplitude is nonzero at \OvertoneSignificance credibility for $\tStart = \tSecondOvertone\, \tMf$ and remains above $3\sigma$ until $\tStart = \OvertoneThreeSigmaTime\,\tMf$.
The significance drops below $1\sigma$ past $\tStart = \tOvertone\, \tMf$, when the data become consistent with a single damped sinusoid.
Additionally, $\tStart = \tSecondOvertone\, \tMf$ emerges as an inflection point in the amplitude trend (gray shading): for later start times, the recovered amplitudes decay consistently with the expected exponential decay.
The divergence between the inferred overtone amplitude and the expected decay for $\tStart < \tSecondOvertone\, \tMf$ hints at unmodeled features in the data, further explored in a forthcoming paper with extended models~\cite{GW250114Companion}.
At $\tStart = \tSecondOvertone\, \tMf$, the two-mode SNR is \OvertoneSNR, and the (redshifted) frequency and damping rate of the overtone are $f_{221} = \OvertoneFreq\, \mathrm{Hz}$ and $\gamma_{221} = \OvertoneDamp\, \mathrm{Hz}$, respectively.
The \eventname post-merger signal after $\tStart = \tSecondOvertone\, \tMf$ has the same SNR as GW150914 in its entirety.

\section{The Kerr Nature of the remnant}

\begin{figure}
\centering
\includegraphics[width=\linewidth]{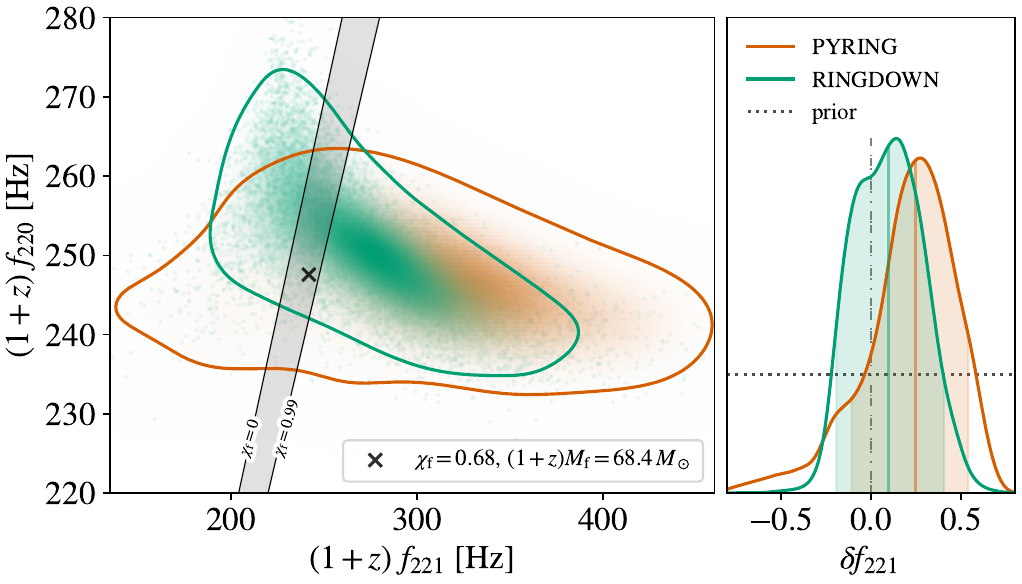}
\caption{Spectroscopic test of the Kerr nature of the remnant black hole for $\tStart=\tSecondOvertone\,\tMf$. 
Left: 90\% posterior for the observed fundamental and overtone frequencies, compared to the range allowed by the Kerr spectrum (black shaded region) for any black hole mass (vertical span) and spin (horizontal span).
The cross marks reference values from the full-signal analysis.
Right: posterior on the deviation \(\delta f_{221}\) of the frequency of the first overtone from the Kerr spectrum, with shading and a line showing the 90\% credible region and the median respectively, a vertical line showing the Kerr prediction of \(\delta f_{221} = 0\), and a horizontal line showing the prior.
The \pyring analysis (orange) starts ${\sim}\PyRingDelay\,\tMf$ after \ringdown (green), which is why the posteriors are not identical, see the Supplement\citesupp.
The observed spectrum is consistent with Kerr to $\Specsign$.
}
\label{fig:221_kerr}
\end{figure}

Identification of two modes in the data can be used to test the Kerr nature of the merger remnant.
Consistency with Kerr amounts to the four observables (the frequency and damping rate of each mode) agreeing with the Kerr spectrum for some mass and spin.
Black hole spectroscopy \cite{Dreyer:2003bv,Gossan:2011ha,Berti:2025hly} started in earnest with searches for $\ell = |m| =2$ modes in GW150914~\cite{LIGOScientific:2016lio,Prix:T1500618,Carullo:2019flw,Isi:2019aib,Isi:2022mhy,Cotesta:2022pci,Isi:2023nif,Finch:2022ynt,Correia:2023bfn,Wang:2023xsy,Ma:2023vvr} and has since been extended to further events and modes \cite{Brito:2018rfr,LIGOScientific:2020tif,LIGOScientific:2021sio,LIGOScientific:2020iuh,LIGOScientific:2020ufj,Capano:2021etf,Siegel:2023lxl,LIGOScientific:2025rsn,GW230814}.
To constrain deviations away from the Kerr spectrum, we enhance our model with two additional parameters that allow for deviations in the frequency and damping rate of the overtone respectively: $f_{221} = f^{\rm (Kerr)}_{221}(\mf, \chif)\, \exp(\df)$ and $\gamma_{221} = \gamma^{\rm (Kerr)}_{221}(\mf, \chif)\,\exp(\dg)$.
The frequencies and damping rates of both modes are now parametrized via the remnant mass and spin, $\mf$ and $\chif$, and the deviations, $\df$ and $\dg$.
The Kerr spectrum is recovered for $\df = \dg = 0$.

With this setup, we constrain deviations  from the Kerr spectrum at $\tStart = \tSecondOvertone\,\tMtot$ to be $\df = \dfConstraint$, as seen in Fig.~\ref{fig:221_kerr}, 
while \dg remains uninformative within its prior.
This bolsters confidence in the identification of this mode as the overtone, and establishes consistency with Kerr frequencies to $\Specsign$.
Results for a variety of start times are presented in a forthcoming paper~\cite{GW250114Companion}.
This is the first constraint of this kind derived from data confidently removed from the signal peak \cite{Isi:2019aib,LIGOScientific:2020tif,LIGOScientific:2021sio}.

\section{Hawking's Area Law}

\emph{Background.} The second law of black hole mechanics, originally proven by \citet{PhysRevLett.26.1344} (but previously explored as irreducible mass by~\citet{PhysRevLett.25.1596, Christodoulou:1971pcn} and stated by~\citet{Penrose_1971}), states that the black hole horizon area cannot decrease in time.
A direct consequence is an upper limit of 50\% on the efficiency of gravitationally radiating processes in systems with initially vanishing binding energy; this is further limited to ${\sim}29\%$ for non spinning black holes \cite{PhysRevLett.26.1344}.

The area law relies on three conditions.
The first is the null-energy condition, a restriction on the properties of matter.
It is violated, for example, by Hawking radiation which extracts energy from a black hole and causes its horizon to shrink~\cite{hawking1974blackhole}---in this case, the area law is superseded by a generalized law that considers both the entropy of the black hole and that of the radiation~\cite{PhysRevD.9.3292, hawking1975particle}.
The second is the premise that the observed objects are black holes and weak cosmic censorship holds, i.e., no naked singularities.
Alternative compact objects~\cite{Mazur:2004fk,Mathur:2005zp,Cardoso:2019rvt,Cayuso:2020lca} have modified entropy and violate the null-energy condition; their interactions could violate the area law.
The third is general relativity; 
the area law can be violated in alternative theories~\cite{Creighton:1995au,Giudice:2016zpa,Faraoni:2010yi,Cardoso:2018ptl,Cayuso:2023xbc}.
Testing the area law thus amounts to testing for physical behavior that violates at least one of these conditions.

In the context of a binary merger, the area law imposes an increase in the horizon area of the remnant with respect to the total area of the initial black holes~\cite{PhysRevLett.26.1344}, providing a testable prediction~\cite{Hughes:2004vw,Cabero:2017avf,Isi:2020tac,Correia:2023ipz,Akyuz:2025seg}.
However, since black hole areas are not direct observables, extracting them hinges on certain assumptions: (i) \eventname originated from a quasicircular merging binary, (ii) general relativity is a good approximation away from highly dynamical regions, and (iii) the black holes are well described by the Kerr metric.
The latter guides the black hole states we probe: the initial black holes are considered at wide separations, while the final black hole is considered in its asymptotic state.
We thus adopt the Kerr area formula~\cite{Bardeen:1973gs},
\begin{equation}
\mathcal{A}(m, \chi)=8 \pi\left(\frac{G m}{c^2}\right)^2\left(1+\sqrt{1-\chi^2}\right)\,,
\label{eq:areaKerr}
\end{equation}
for a black hole of mass $m$ and spin $\chi$. 
This need not hold beyond vacuum general relativity~\cite{Kleihaus:2015aje,Cano:2019ore,Cardoso:2018ptl,Cayuso:2023xbc}.

\emph{Method.} We extract the properties of the initial and final black holes independently from the pre- and post-merger signal respectively, discarding data in between.
Our test, therefore, probes for violations during the most nonlinear and dynamical portion of the signal, which it excludes.
It is sensitive to any nonstandard process that may alter the radiated energy or momentum,
or any physics that modifies the ringdown spectrum sufficiently to bias the inferred remnant mass and spin, e.g., electromagnetic charge or other deviations from Kerr.
Most generally, it is sensitive to physics that breaks any of the premises above and leads the inspiral and ringdown regimes to be better described independently than coherently \cite{Isi:2020tac}.
Comparing the inspiral to the ringdown in this fashion can provide complementary constraints to those derived from the ringdown alone \cite{Carullo:2021oxn,Gu:2023eaa,Carullo:2021dui,Silva:2022srr,Maenaut:2024oci,Crescimbeni:2024sam,Chung:2025wbg}.
We truncate the data in the time domain \cite{Cabero:2017avf,Isi:2020tac,Isi:2021iql,Miller:2023ncs,Miller:2025eak} rather than the frequency domain \cite{Ghosh:2016qgn,Ghosh:2017gfp}, as there is no exact one-to-one correspondence between signal time and Fourier frequency beyond the adiabatic inspiral~\cite{PhysRevD.44.3819,Droz:1999qx}.

\emph{The initial horizon area.} The initial black holes are considered at wide separations, where they obey the Kerr metric and the total area is the sum of the individual areas.
The quantities reported in data analysis differ by the amount of energy and angular momentum absorbed by the black holes throughout the binary evolution, which is negligible for comparable-mass systems~\cite{Poisson:2004cw,Scheel:2014ina}.
We therefore infer the black hole properties from the inspiral signal and interpret them directly as the infinite-separation quantities.
Assuming that general relativity describes the inspiral~\cite{Will:2014kxa,LIGOScientific:2016lio,Yunes:2016jcc,LIGOScientific:2020tif,LIGOScientific:2019fpa,LIGOScientific:2021sio}
and the binary orbit is quasicircular~\cite{Divyajyoti:2023rht}, we model the signal with \NRSur, up to a preselected time, $\tEnd$, that is quoted in units of the total mass from the full-signal analysis, $\tMtot\equiv\left(1+z\right) G M /c^3$=\,\tMRef.
We then infer (among other parameters) the masses and spins of the initial black holes, informed only by data before $\tEnd$ using the \TDInf inference package \cite{Isi:2020tac,Miller:2023ncs,Miller:2025eak}, and from them the initial area $\mathcal{A}_i=\mathcal{A}_1+\mathcal{A}_2$.
See the Supplement for details\citesupp.

\begin{figure}
\centering
\includegraphics[width=\linewidth]{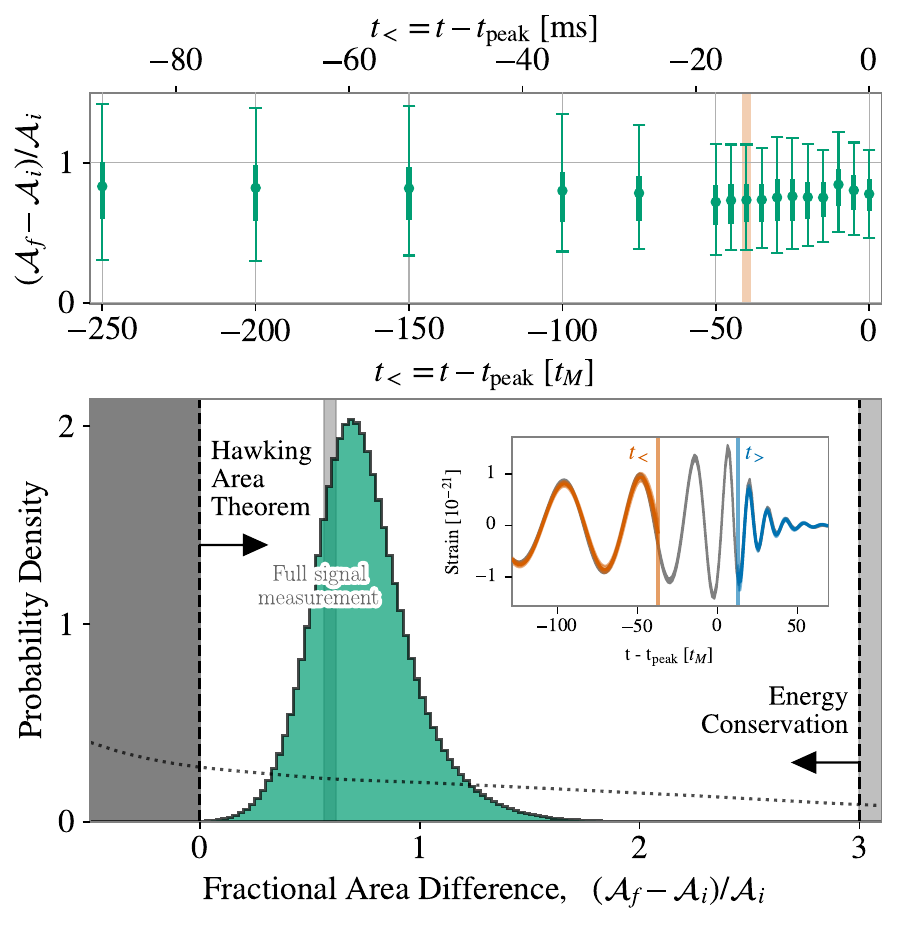}
\caption{Fractional difference between the area of the final black hole, $\mathcal{A}_f$, and the initial black holes, $\mathcal{A}_i$.
Top: Posterior as a function of the pre-merger truncation time, $\tEnd=t-\tRefSymbol$. 
Bars represent 90\% (thin line) and 50\% (thick line) credible intervals around the median (circle). 
The vertical orange band denotes $\tEnd=\Pretime\,\tMtot$, when the gravitational-wave luminosity is at $10\%$ of its maximum.
Bottom: Posterior (histogram) and prior (dotted line) for $\tEnd=\Pretime\,\tMtot$.
The histogram is produced from all pairwise combinations of \ringdown and \pyring post-merger samples with \TDInf pre-merger samples.
The shading on the right highlights configurations that would violate energy conservation $M_f\leq m_1+m_2$, a bound saturated for maximally-spinning initial black holes that merge into a nonspinning remnant \cite{1962RSPSA.269...21B,PhysRevLett.26.1344,1962RSPSA.270..103S}.
The shading on the left highlights configurations that would violate the area law, $(\mathcal{A}_f - \mathcal{A}_i)/\mathcal{A}_i<0$. 
The vertical gray band is the 90\% credible interval from the full-signal analysis used in Fig.~\ref{fig:PE_mass}, i.e., the general relativity expectation for \eventname.
The inset illustrates the analysis via the signal reconstruction in LIGO Livingston from the full-signal (gray), and when split into the pre-merger (orange) and post-merger (blue) stages;
the merger phase is included in neither analysis.
We find that $\mathcal{A}_f > \mathcal{A}_i$ to high credibility, indicating that \eventname obeys Hawking's area law.
}
\label{fig:arealaw}
\end{figure}

To interpret the pre-merger truncation times, we use the gravitational-wave luminosity to proxy how relativistic the system is~\cite{Blanchet:2013haa}.
Based on the full-signal analysis, in the Supplement we show that the luminosity is sharply peaked at merger and drops to $10\%$ of its maximum at $\FluxTenPercent\,\tMtot$ before $\tRefSymbol$\citesupp.
We present results for an array of $\tEnd$ in Fig.~\ref{fig:arealaw} and further highlight $\tEnd=\Pretime\,\tMtot$, a choice that excludes the two loudest signal cycles.
For reference, the total signal SNR before $\tEnd=\Pretime\,\tMtot$ is $\PreSNR$; even truncated, \eventname has a higher SNR than any other signal to date~\cite{GWTC4results}.

\emph{The final horizon area.} The mass and spin of the remnant are inferred through the post-merger signal.
Picking an analysis start time requires balancing: (i) independence from the full-signal analysis that assumes the area law, (ii) maximizing the amount of data and thus the SNR, and (iii) using a model within its regime of validity.
Since modes are not orthogonal, the more complex two-mode model is not universally preferable, as the overtone may degrade the inference especially if it is not detectable.
Therefore we adopt the most parsimonious model (single $\ell = |m| =2, n=0$ mode) at its earliest time of applicability (conservatively, $\tStart= \tOvertone\,\tMf$ when the overtone significance is ${<}1\sigma$) thus minimizing potential contamination~\cite{Kastha:2021chr}.
We then use the mode's inferred frequency and damping rate to calculate the mass and spin of the final Kerr black hole and hence its area, $\mathcal{A}_f$.

\emph{Results.} Figure~\ref{fig:arealaw} shows the fractional difference between the final and initial areas, $(\mathcal{A}_f-\mathcal{A}_i)/\mathcal{A}_i$, where cosmological redshift factors cancel out
because the area is proportional to the mass squared.
For $\tEnd=\Pretime\,\tMtot$, $\mathcal{A}_f > \mathcal{A}_i$ at the $\ALsign$ level, computed as described in the Supplement\citesupp.
We find consistency with an increase in the area to at least $\ALsignminimum$ for all times $\tEnd > -250\,\tMtot$, exceeding $5\sigma$ for $\tEnd \geq \AreaLawFiveSigmaTime\,\tMtot$.
In the Supplement we present results for more times and models\citesupp; the two-mode model at its earliest time of applicability, $\tStart=\tSecondOvertone\,\tMf$, also agrees with the area law at $\ALsignOvertone\sigma$.
All results are further consistent with the general relativity expectation for \eventname, obtained from the full-signal analysis with \NRSur used in Fig.~\ref{fig:PE_mass}, which considers the entire signal coherently and obeys the area law a priori.
The strength of \eventname enables such tests even while excluding the loudest portions of the signal and without any modeling of the nonlinear merger dynamics through numerical relativity, cf.~inset of Fig.~\ref{fig:arealaw}.
Similar analyses of GW150914~\cite{Isi:2020tac,Correia:2023ipz} yielded ${\sim}2\sigma$ results, albeit with less conservative assumptions: a pre-merger analysis that extended to the waveform peak and a quasinormal mode model that presumed circular polarization.

\section{Conclusion}

The gravitational-wave signal \eventname is a milestone in the decade-long history of gravitational-wave science. 
With its high SNR, \eventname offers an exquisitely clear view of the highly dynamical process by which two black holes merge to give rise to a remnant black hole. 
Data from LIGO Hanford and LIGO Livingston are consistent with multiple quasinormal modes of a remnant Kerr black hole and with Hawking's area law.
Further precision tests of general relativity and the ringdown are presented in a forthcoming publication~\cite{GW250114Companion}.
Our results suggest that astrophysical black holes are indeed extremely simple objects that follow general relativity and the Kerr description.
The strongly perturbed merger remnant settles into a higher-entropy, quiescent state within a few dynamical timescales.
The next decade of gravitational-wave science is bound to enhance our view of these highly dynamical, relativistic systems.

Strain data from the LIGO detectors for
\eventname are available from the Gravitational Wave
Open Science Center \cite{GWOSC_page}. All the material required for reproducing the figures, including scripts and posterior distributions from the analyses, is available
in the data release \cite{zenodo_data_release}.

\section{Acknowledgments}

This work made use of the following software, listed in alphabetical order:
\arviz \cite{arviz_2019}, %
\Asimov \cite{Williams:2022pgn},
\astropy \cite{astropy:2013,astropy:2018,astropy:2022}, %
\BayesWave \cite{Cornish:2020dwh},
\bilby \cite{Ashton:2018jfp, Romero-Shaw:2020owr}, 
\cmasher \cite{vanderVelden2020},
\cpnest \cite{cpnest},
\CWB \cite{Klimenko:2015ypf,Mishra:2024zzs},
\dynesty \cite{Speagle:2019ivv},
\emcee \cite{ForemanMackey:2013},
\GSTLAL \cite{Cannon:2011rj, Messick:2016aqy, Sachdev:2019vvd, Hanna:2019ezx, Cannon:2020qnf, Tsukada:2023edh, Ewing:2023qqe, Sakon:2022ibh, Joshi:2025nty, Joshi:2025zdu, Joshi:2023ltf, Ray:2023nhx}, 
\gwpy \cite{gwpy, duncan_macleod_2024_12734623}, %
\hfivepy \cite{collette_python_hdf5_2014}, %
\Phenom \citep{Pratten:2020ceb, Colleoni:2024knd}, 
\PhenomOfour \citep{Hamilton:2021pkf, Thompson:2023ase},
\jax \cite{jax2018github}, %
\jupyter and \ipython \cite{ipython_2007,jupyter_2016,jupyter_2021}, %
\lalsuite \cite{lalsuite}, %
\swiglal \cite{swiglal}, %
\matplotlib \cite{matplotlib_3_7_3_zenodo,Hunter:2007}, %
\MBTA \cite{Aubin:2020goo,Allene:2025saz}, 
\MLy \cite{Skliris:2020qax}, 
\NRSur \citep{Varma:2019csw},
\numpy \cite{numpy}, %
\numpyro \citep{phan2019composable, bingham2019pyro}, %
\pandas \citep{mckinney-proc-scipy-2010, pandas_13819579}, %
\pesummary \cite{Hoy:2020vys}, %
\PYCBC \cite{DalCanton:2020vpm}, 
\pyring \cite{pyRing, Carullo:2019flw}, 
\pySEOBNR \cite{Mihaylov:2023bkc},
\python \cite{python},
\qnm \cite{qnm, qnm_joss},
\rift~\cite{Pankow:2015cra,Lange:2017wki,Wysocki:2019grj},
\ringdown \cite{ringdown,Isi:2021iql}, 
\scipy \citep{2020SciPy-NMeth, scipy_10543017}, %
\seaborn \citep{Waskom2021}, %
\EOB \citep{Pompili:2023tna, Khalil:2023kep, vandeMeent:2023ols, Ramos-Buades:2023ehm}, 
\EOBEcc \citep{Gamboa:2024hli},
\TEOBEcc \citep{Nagar:2024dzj},
\SPIIR \cite{Chu:2020pjv}, 
\TDInf \cite{TDInf, Miller:2023ncs,Miller:2025eak}, 
\tqdm \citep{tqdm_14002015}, %
Software citation information was aggregated using \texttt{\href{https://www.tomwagg.com/software-citation-station/}{The Software Citation Station}} \citep{software-citation-station-paper, software-citation-station-zenodo}.

\ifprintauthors
This material is based upon work supported by NSF's LIGO Laboratory, which is a
major facility fully funded by the National Science Foundation.
The authors also gratefully acknowledge the support of
the Science and Technology Facilities Council (STFC) of the
United Kingdom, the Max-Planck-Society (MPS), and the State of
Niedersachsen/Germany for support of the construction of Advanced LIGO 
and construction and operation of the GEO\,600 detector. 
Additional support for Advanced LIGO was provided by the Australian Research Council.
The authors gratefully acknowledge the Italian Istituto Nazionale di Fisica Nucleare (INFN),  
the French Centre National de la Recherche Scientifique (CNRS) and
the Netherlands Organization for Scientific Research (NWO)
for the construction and operation of the Virgo detector
and the creation and support  of the EGO consortium. 
The authors also gratefully acknowledge research support from these agencies as well as by 
the Council of Scientific and Industrial Research of India, 
the Department of Science and Technology, India,
the Science \& Engineering Research Board (SERB), India,
the Ministry of Human Resource Development, India,
the Spanish Agencia Estatal de Investigaci\'on (AEI),
the Spanish Ministerio de Ciencia, Innovaci\'on y Universidades,
the European Union NextGenerationEU/PRTR (PRTR-C17.I1),
the ICSC - CentroNazionale di Ricerca in High Performance Computing, Big Data
and Quantum Computing, funded by the European Union NextGenerationEU,
the Comunitat Auton\`oma de les Illes Balears through the Conselleria d'Educaci\'o i Universitats,
the Conselleria d'Innovaci\'o, Universitats, Ci\`encia i Societat Digital de la Generalitat Valenciana and
the CERCA Programme Generalitat de Catalunya, Spain,
the Polish National Agency for Academic Exchange,
the National Science Centre of Poland and the European Union - European Regional
Development Fund;
the Foundation for Polish Science (FNP),
the Polish Ministry of Science and Higher Education,
the Swiss National Science Foundation (SNSF),
the Russian Science Foundation,
the European Commission,
the European Social Funds (ESF),
the European Regional Development Funds (ERDF),
the Royal Society, 
the Scottish Funding Council, 
the Scottish Universities Physics Alliance, 
the Hungarian Scientific Research Fund (OTKA),
the French Lyon Institute of Origins (LIO),
the Belgian Fonds de la Recherche Scientifique (FRS-FNRS), 
Actions de Recherche Concert\'ees (ARC) and
Fonds Wetenschappelijk Onderzoek - Vlaanderen (FWO), Belgium,
the Paris \^{I}le-de-France Region, 
the National Research, Development and Innovation Office of Hungary (NKFIH), 
the National Research Foundation of Korea,
the Natural Sciences and Engineering Research Council of Canada (NSERC),
the Canadian Foundation for Innovation (CFI),
the Brazilian Ministry of Science, Technology, and Innovations,
the International Center for Theoretical Physics South American Institute for Fundamental Research (ICTP-SAIFR), 
the Research Grants Council of Hong Kong,
the National Natural Science Foundation of China (NSFC),
the Israel Science Foundation (ISF),
the US-Israel Binational Science Fund (BSF),
the Leverhulme Trust, 
the Research Corporation,
the National Science and Technology Council (NSTC), Taiwan,
the United States Department of Energy,
and
the Kavli Foundation.
The authors gratefully acknowledge the support of the NSF, STFC, INFN and CNRS for provision of computational resources.

This work was supported by MEXT,
the JSPS Leading-edge Research Infrastructure Program,
JSPS Grant-in-Aid for Specially Promoted Research 26000005,
JSPS Grant-in-Aid for Scientific Research on Innovative Areas 2402: 24103006,
24103005, and 2905: JP17H06358, JP17H06361 and JP17H06364,
JSPS Core-to-Core Program A.\ Advanced Research Networks,
JSPS Grants-in-Aid for Scientific Research (S) 17H06133 and 20H05639,
JSPS Grant-in-Aid for Transformative Research Areas (A) 20A203: JP20H05854,
the joint research program of the Institute for Cosmic Ray Research,
University of Tokyo,
the National Research Foundation (NRF),
the Computing Infrastructure Project of the Global Science experimental Data hub
Center (GSDC) at KISTI,
the Korea Astronomy and Space Science Institute (KASI),
the Ministry of Science and ICT (MSIT) in Korea,
Academia Sinica (AS),
the AS Grid Center (ASGC) and the National Science and Technology Council (NSTC)
in Taiwan under grants including the Science Vanguard Research Program,
the Advanced Technology Center (ATC) of NAOJ,
and the Mechanical Engineering Center of KEK.

Additional acknowledgements for support of individual authors may be found in the following document: \\
\texttt{https://dcc.ligo.org/LIGO-M2300033/public}.
For the purpose of open access, the authors have applied a Creative Commons Attribution (CC BY)
license to any Author Accepted Manuscript version arising.
We request that citations to this article use `A. G. Abac {\it et al.} (LIGO-Virgo-KAGRA Collaboration), ...' or similar phrasing, depending on journal convention.

 \fi

\bibliography{source/bibliography}

\ifprintsupplement
\clearpage
\appendix
\onecolumngrid
\section{Supplement to ``\PaperTitle''}
\section{I. Observation and inference details}
\label{sec:parameter-estimation}

Here we describe additional details relevant to the detection and parameter inference of \eventname.
See also Ref.~\cite{GWTC4methods} for more details on the methods for identifying and characterizing gravitational-wave transients.

The \GSTLAL~\cite{Cannon:2011rj, Messick:2016aqy, Sachdev:2019vvd, Hanna:2019ezx, Cannon:2020qnf, Tsukada:2023edh, Ewing:2023qqe, Sakon:2022ibh, Joshi:2025nty, Joshi:2025zdu, Joshi:2023ltf, Ray:2023nhx} search pipeline was the first to report that it had identified \eventname with high confidence \GSTLALTimeDelay seconds after the signal arrived.
Within $\TimeDelay\,$s of the signal crossing the Earth, all operating low-latency search pipelines reported detections, including those targeting unmodelled transient events: \MLy\cite{Skliris:2020qax}, \SPIIR~\cite{Chu:2020pjv}, \MBTA~\cite{Aubin:2020goo,Allene:2025saz}, \PYCBC~\cite{DalCanton:2020vpm}, and \CWB~\cite{Klimenko:2015ypf,Mishra:2024zzs}.
All pipelines assigned a false alarm rate of \FARThreshold years or lower.
The modeled pipelines reported network matched-filter SNRs~\cite{Allen:2005fk} ranging from \networkMatchedFilterSNRlower to \networkMatchedFilterSNRupper. 
The variation is due to the discreteness in the sets of filter waveforms used~\cite{GWTC4methods} as well as differences in the estimation of the detector noise power spectral density at that time.
This signal has by far the highest SNR yet recorded, with the previous highest being the single-detector signal GW230814\_230901 with a matched-filter SNR of \SNRTwentyThreeZeroEightNineteen~\cite{GW230814,GWTC4results}.
The two-detector sky localization was released publicly at \SkymapTime, $\SkymapTimeDelay\,$s after the signal reached the detectors. 
Based on the classification reported by the search pipelines~\cite{LVK:2025gcn38932}, \eventname was determined to have a greater than \PBlackHoleLowerBound probability of having a binary black hole origin. 
The data-quality report noted no issues in LIGO Hanford. 
Some excess power was reported in LIGO Livingston, although glitch subtraction~\cite{Cornish:2020dwh,Hourihane:2022doe,Hourihane:2025vxc} was not needed due to the power being confined to low frequencies.

We infer the source properties following standard procedures~\cite{GWTC4methods} using the \Asimov workflow package \citep{Williams:2022pgn}.
We analyze $\duration\,$s of data in the $\fmin{-}\fmax\,$Hz frequency range.
The detector noise power spectral density is obtained with \BayesWave \citep{Littenberg:2014oda,Cornish:2020dwh}. 
We use the inference library \bilby \citep{Ashton:2018jfp, Romero-Shaw:2020owr} and the \dynesty nested sampler \citep{Speagle:2019ivv} to sample from the posterior distribution of the source parameters for the main analyses that assume quasicircular orbits.
We further use \bilby and \rift~\cite{Pankow:2015cra,Lange:2017wki,Wysocki:2019grj} for analyses that allow for eccentric orbits, but are restricted to aligned spins. 
We marginalize over the luminosity distance with a prior that corresponds to a uniform merger rate in co-moving volume using the Planck 2015 $\Lambda$CDM cosmology model \citep{Planck:2015fie}. 
We further marginalize over uncertainties in the detector calibration; the $1\sigma$ bounds on the calibration uncertainty remain within \AmpCalib in the amplitude and \PhaseCalib degrees in phase for both detectors within the analysis band.

We model the signal with four waveform models that include the effects of spin-precession and higher-order radiation modes but are restricted to quasi-circular orbits: \NRSur \citep{Varma:2019csw}, \Phenom \citep{Pratten:2020ceb, Colleoni:2024knd}, \EOB \citep{Pompili:2023tna, Khalil:2023kep, vandeMeent:2023ols, Ramos-Buades:2023ehm}, and \PhenomOfour \citep{Hamilton:2021pkf, Thompson:2023ase}.
Analysis settings for \Phenom, \EOB, and \PhenomOfour ensure that modes up to $\ell=4$ are fully accounted for.
Due to the finite length of \NRSur, the $\ell=3$ and $\ell=4$ modes enter above $\EllThreeFrequency\,$Hz and $\EllFourFrequency\,$Hz respectively.  
Priors are uniform in detector-frame component masses and spin magnitudes, isotropic in spin orientations, isotropic in the binary's orientation, uniform in merger time and coalescence phase, uniform in comoving volume, and isotropic in sky location. 
Figure~\ref{fig:PE_waveforms} shows the two-dimensional posterior for the component masses and effective inspiral and precessing spin parameters obtained from each model.
Even though the posteriors are not identical, they all lead to a similar interpretation of \eventname as an equal-mass system with small spins. 
We list the 90\% credible intervals for further parameters in Table~\ref{table:individual_pe}.

We additionally consider two models that allow for eccentric orbits but are restricted to aligned spins: \EOBEcc~\cite{Gamboa:2024hli} and \TEOBEcc~\cite{Nagar:2024dzj}.
We use identical settings to the main runs, and adopt uniform priors for the orbital eccentricity and relativistic (mean) anomaly for \EOBEcc (\TEOBEcc).
Both analyses yield no evidence for eccentricity, with 90\% upper limits of $e \EccUpperLimit$ at an orbit-averaged gravitational-wave frequency~\cite{Ramos-Buades:2023yhy} of $\EccRefFreq~$Hz.

\begin{table}
\begin{ruledtabular}
    \caption{Source properties of \eventname for various parameters and four waveform models for spin-precessing, quasicircular systems. We report the median values together with the 90\% symmetric credible intervals at a reference frequency of $20\, 	\text{Hz}$. For parameters that rail against the minimum (maximum) possible values we display upper (lower) limits at the 90\% credible level.
    }
    \label{table:individual_pe}
    \renewcommand{\arraystretch}{1.2}
    \begin{center}
    \begin{tabular}{l r r r r r}
    Parameter & \NRSur & \EOB & \Phenom & \PhenomOfour \\
    \hline
            Primary mass $m_1 / \Msun$ & $\massonesourcemed{GW250114_NRSUR}^{+\massonesourceplus{GW250114_NRSUR}}_{-\massonesourceminus{GW250114_NRSUR}}$ & $\massonesourcemed{GW250114_SEOB}^{+\massonesourceplus{GW250114_SEOB}}_{-\massonesourceminus{GW250114_SEOB}}$ & $\massonesourcemed{GW250114_Phenom}^{+\massonesourceplus{GW250114_Phenom}}_{-\massonesourceminus{GW250114_Phenom}}$ & $\massonesourcemed{GW250114_XO4a}^{+\massonesourceplus{GW250114_XO4a}}_{-\massonesourceminus{GW250114_XO4a}}$ \\
        Secondary mass $m_2 / \Msun $ & $\masstwosourcemed{GW250114_NRSUR}^{+\masstwosourceplus{GW250114_NRSUR}}_{-\masstwosourceminus{GW250114_NRSUR}}$ & $\masstwosourcemed{GW250114_SEOB}^{+\masstwosourceplus{GW250114_SEOB}}_{-\masstwosourceminus{GW250114_SEOB}}$ & $\masstwosourcemed{GW250114_Phenom}^{+\masstwosourceplus{GW250114_Phenom}}_{-\masstwosourceminus{GW250114_Phenom}}$ & $\masstwosourcemed{GW250114_XO4a}^{+\masstwosourceplus{GW250114_XO4a}}_{-\masstwosourceminus{GW250114_XO4a}}$ \\
        Mass ratio $q = m_2 / m_1$ & $\geq \massratiotenpercent{GW250114_NRSUR}$& $\geq \massratiotenpercent{GW250114_SEOB}$& $\geq \massratiotenpercent{GW250114_Phenom}$& $\geq \massratiotenpercent{GW250114_XO4a}$\\
        Total mass $M / \Msun$ & $\totalmasssourcemed{GW250114_NRSUR}^{+\totalmasssourceplus{GW250114_NRSUR}}_{-\totalmasssourceminus{GW250114_NRSUR}}$ & $\totalmasssourcemed{GW250114_SEOB}^{+\totalmasssourceplus{GW250114_SEOB}}_{-\totalmasssourceminus{GW250114_SEOB}}$ & $\totalmasssourcemed{GW250114_Phenom}^{+\totalmasssourceplus{GW250114_Phenom}}_{-\totalmasssourceminus{GW250114_Phenom}}$ & $\totalmasssourcemed{GW250114_XO4a}^{+\totalmasssourceplus{GW250114_XO4a}}_{-\totalmasssourceminus{GW250114_XO4a}}$ \\
        Detector-frame total mass $(1+z) \ M/M_\odot$ & $\totalmassmed{GW250114_NRSUR}^{+\totalmassplus{GW250114_NRSUR}}_{-\totalmassminus{GW250114_NRSUR}}$ & $\totalmassmed{GW250114_SEOB}^{+\totalmassplus{GW250114_SEOB}}_{-\totalmassminus{GW250114_SEOB}}$ & $\totalmassmed{GW250114_Phenom}^{+\totalmassplus{GW250114_Phenom}}_{-\totalmassminus{GW250114_Phenom}}$ & $\totalmassmed{GW250114_XO4a}^{+\totalmassplus{GW250114_XO4a}}_{-\totalmassminus{GW250114_XO4a}}$ \\
        Chirp mass $\mathcal{M}/M_\odot$  & $\chirpmasssourcemed{GW250114_NRSUR}^{+\chirpmasssourceplus{GW250114_NRSUR}}_{-\chirpmasssourceminus{GW250114_NRSUR}}$ & $\chirpmasssourcemed{GW250114_SEOB}^{+\chirpmasssourceplus{GW250114_SEOB}}_{-\chirpmasssourceminus{GW250114_SEOB}}$ & $\chirpmasssourcemed{GW250114_Phenom}^{+\chirpmasssourceplus{GW250114_Phenom}}_{-\chirpmasssourceminus{GW250114_Phenom}}$ & $\chirpmasssourcemed{GW250114_XO4a}^{+\chirpmasssourceplus{GW250114_XO4a}}_{-\chirpmasssourceminus{GW250114_XO4a}}$ \\
        Detector-frame chirp mass $(1+z)\ \mathcal{M}/M_\odot$ & $\chirpmassmed{GW250114_NRSUR}^{+\chirpmassplus{GW250114_NRSUR}}_{-\chirpmassminus{GW250114_NRSUR}}$ & $\chirpmassmed{GW250114_SEOB}^{+\chirpmassplus{GW250114_SEOB}}_{-\chirpmassminus{GW250114_SEOB}}$ & $\chirpmassmed{GW250114_Phenom}^{+\chirpmassplus{GW250114_Phenom}}_{-\chirpmassminus{GW250114_Phenom}}$ & $\chirpmassmed{GW250114_XO4a}^{+\chirpmassplus{GW250114_XO4a}}_{-\chirpmassminus{GW250114_XO4a}}$ \\
        Final mass $M_{\rm{f}} / \Msun$ & $\finalmasssourcemed{GW250114_NRSUR}^{+\finalmasssourceplus{GW250114_NRSUR}}_{-\finalmasssourceminus{GW250114_NRSUR}}$ & $\finalmasssourcemed{GW250114_SEOB}^{+\finalmasssourceplus{GW250114_SEOB}}_{-\finalmasssourceminus{GW250114_SEOB}}$ & $\finalmasssourcemed{GW250114_Phenom}^{+\finalmasssourceplus{GW250114_Phenom}}_{-\finalmasssourceminus{GW250114_Phenom}}$ & $\finalmasssourcemed{GW250114_XO4a}^{+\finalmasssourceplus{GW250114_XO4a}}_{-\finalmasssourceminus{GW250114_XO4a}}$ \\
        Detector-frame final mass $(1+z) \ M_f/M_\odot$ & $\finalmassmed{GW250114_NRSUR}^{+\finalmassplus{GW250114_NRSUR}}_{-\finalmassminus{GW250114_NRSUR}}$ & $\finalmassmed{GW250114_SEOB}^{+\finalmassplus{GW250114_SEOB}}_{-\finalmassminus{GW250114_SEOB}}$ & $\finalmassmed{GW250114_Phenom}^{+\finalmassplus{GW250114_Phenom}}_{-\finalmassminus{GW250114_Phenom}}$ & $\finalmassmed{GW250114_XO4a}^{+\finalmassplus{GW250114_XO4a}}_{-\finalmassminus{GW250114_XO4a}}$ \\
        Primary spin magnitude $\chi_1$ & $\leq \aoneninetypercent{GW250114_NRSUR}$& $\leq \aoneninetypercent{GW250114_SEOB}$& $\leq \aoneninetypercent{GW250114_Phenom}$& $\leq \aoneninetypercent{GW250114_XO4a}$\\
        Secondary spin magnitude $\chi_2$ & $\leq \atwoninetypercent{GW250114_NRSUR}$& $\leq \atwoninetypercent{GW250114_SEOB}$& $\leq \atwoninetypercent{GW250114_Phenom}$& $\leq \atwoninetypercent{GW250114_XO4a}$\\
        Effective inspiral-spin $\chi_{\rm eff}$ & $\chieffmed{GW250114_NRSUR}^{+\chieffplus{GW250114_NRSUR}}_{-\chieffminus{GW250114_NRSUR}}$ & $\chieffmed{GW250114_SEOB}^{+\chieffplus{GW250114_SEOB}}_{-\chieffminus{GW250114_SEOB}}$ & $\chieffmed{GW250114_Phenom}^{+\chieffplus{GW250114_Phenom}}_{-\chieffminus{GW250114_Phenom}}$ & $\chieffmed{GW250114_XO4a}^{+\chieffplus{GW250114_XO4a}}_{-\chieffminus{GW250114_XO4a}}$ \\
        Effective precessing-spin $\chi_{\rm p}$ & $\chipmed{GW250114_NRSUR}^{+\chipplus{GW250114_NRSUR}}_{-\chipminus{GW250114_NRSUR}}$ & $\chipmed{GW250114_SEOB}^{+\chipplus{GW250114_SEOB}}_{-\chipminus{GW250114_SEOB}}$ & $\chipmed{GW250114_Phenom}^{+\chipplus{GW250114_Phenom}}_{-\chipminus{GW250114_Phenom}}$ & $\chipmed{GW250114_XO4a}^{+\chipplus{GW250114_XO4a}}_{-\chipminus{GW250114_XO4a}}$ \\
        Final spin $\chi_{\rm f}$ & $\finalspinmed{GW250114_NRSUR}^{+\finalspinplus{GW250114_NRSUR}}_{-\finalspinminus{GW250114_NRSUR}}$ & $\finalspinmed{GW250114_SEOB}^{+\finalspinplus{GW250114_SEOB}}_{-\finalspinminus{GW250114_SEOB}}$ & $\finalspinmed{GW250114_Phenom}^{+\finalspinplus{GW250114_Phenom}}_{-\finalspinminus{GW250114_Phenom}}$ & $\finalspinmed{GW250114_XO4a}^{+\finalspinplus{GW250114_XO4a}}_{-\finalspinminus{GW250114_XO4a}}$ \\
        Luminosity distance $D_{\rm L} / \rm{Mpc}$ & $\luminositydistancemed{GW250114_NRSUR}^{+\luminositydistanceplus{GW250114_NRSUR}}_{-\luminositydistanceminus{GW250114_NRSUR}}$ & $\luminositydistancemed{GW250114_SEOB}^{+\luminositydistanceplus{GW250114_SEOB}}_{-\luminositydistanceminus{GW250114_SEOB}}$ & $\luminositydistancemed{GW250114_Phenom}^{+\luminositydistanceplus{GW250114_Phenom}}_{-\luminositydistanceminus{GW250114_Phenom}}$ & $\luminositydistancemed{GW250114_XO4a}^{+\luminositydistanceplus{GW250114_XO4a}}_{-\luminositydistanceminus{GW250114_XO4a}}$ \\
        Viewing angle $\Theta / \rm{rad}$ & $\thetajnmed{GW250114_NRSUR}^{+\thetajnplus{GW250114_NRSUR}}_{-\thetajnminus{GW250114_NRSUR}}$ & $\thetajnmed{GW250114_SEOB}^{+\thetajnplus{GW250114_SEOB}}_{-\thetajnminus{GW250114_SEOB}}$ & $\thetajnmed{GW250114_Phenom}^{+\thetajnplus{GW250114_Phenom}}_{-\thetajnminus{GW250114_Phenom}}$ & $\thetajnmed{GW250114_XO4a}^{+\thetajnplus{GW250114_XO4a}}_{-\thetajnminus{GW250114_XO4a}}$ \\
        Source redshift $z$ & $\redshiftmed{GW250114_NRSUR}^{+\redshiftplus{GW250114_NRSUR}}_{-\redshiftminus{GW250114_NRSUR}}$ & $\redshiftmed{GW250114_SEOB}^{+\redshiftplus{GW250114_SEOB}}_{-\redshiftminus{GW250114_SEOB}}$ & $\redshiftmed{GW250114_Phenom}^{+\redshiftplus{GW250114_Phenom}}_{-\redshiftminus{GW250114_Phenom}}$ & $\redshiftmed{GW250114_XO4a}^{+\redshiftplus{GW250114_XO4a}}_{-\redshiftminus{GW250114_XO4a}}$ \\
        Network $(\ell=4,|m|=4)$ mode SNR $\rho_{44}$ & $\snrfourfourmultipolemed{GW250114_NRSUR}^{+\snrfourfourmultipoleplus{GW250114_NRSUR}}_{-\snrfourfourmultipoleminus{GW250114_NRSUR}}$ & $\snrfourfourmultipolemed{GW250114_SEOB}^{+\snrfourfourmultipoleplus{GW250114_SEOB}}_{-\snrfourfourmultipoleminus{GW250114_SEOB}}$ & $\snrfourfourmultipolemed{GW250114_Phenom}^{+\snrfourfourmultipoleplus{GW250114_Phenom}}_{-\snrfourfourmultipoleminus{GW250114_Phenom}}$ & $\snrfourfourmultipolemed{GW250114_XO4a}^{+\snrfourfourmultipoleplus{GW250114_XO4a}}_{-\snrfourfourmultipoleminus{GW250114_XO4a}}$ \\
    \end{tabular}
    \end{center}
\end{ruledtabular}
\end{table}
\begin{figure}
    \centering
    \includegraphics[width=0.45\linewidth, trim={0 0.25cm 0.02cm 0},clip]{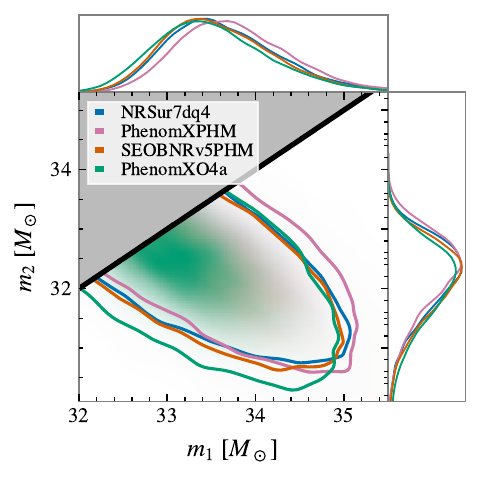}
    \includegraphics[width=0.45\linewidth, trim={0.2cm 0.1cm 0.2cm 0.2},clip]{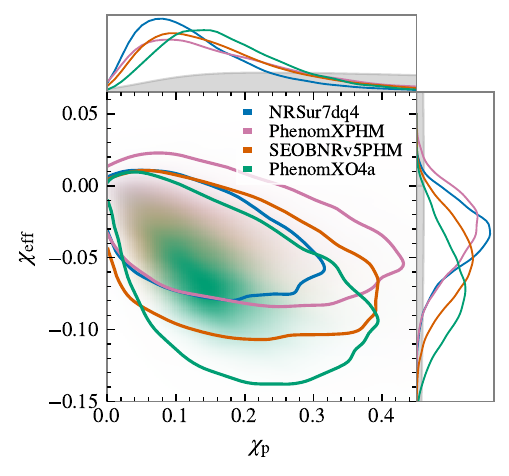}
    \caption{Marginal posterior distributions for the source-frame component masses (left) and effective inspiral and precessing spin parameters (right) of \eventname using four waveform models for spin-precessing, quasicircular systems.
    All models favor an equal-mass system with small spins.}
    \label{fig:PE_waveforms}
\end{figure}

\section{II. Technical details and further results for the post-merger analysis}
\label{app:post-merger}
\subsection{IIA. Reference parameters}

The ringdown analyses in the main text are set up based on a number of reference parameters derived from a preliminary full-signal \NRSur analysis, which are also consistent with the final production analysis.
The first input needed is some guidance on the inferred merger time.
As a proxy for this, we adopt the inferred peak of the gravitational-wave strain, integrated over the celestial sphere around the source; this quantity is independent of the observer's orientation and is commonly adopted as a reference in numerical-relativity studies \cite[e.g.,][]{Mitman:2025hgy,Gao:2025zvl}.
This is also the quantity adopted by the \NRSur approximant as the definition of the coalescence time \cite{Varma:2019csw}.
For a waveform decomposed in terms of spin-weighted spherical harmonics ${}_{-2} Y_{\ell m}(\theta, \phi)$ such that
\begin{equation}
    h_+ - i h_\times = \sum_{\ell m} h_{\ell m}(t)\, {}_{-2} Y_{\ell m}(\theta, \phi)\,,
\end{equation}
the peak of the strain over the celestial sphere is given by
\begin{equation} \label{eq:peak-time}
    t_{\text{peak}} = \mathrm{argmax}_{t}\left[\sum_{\ell m} \left|h_{\ell m}(t)\right|^2\right] \, ,
\end{equation}
which is manifestly invariant under rotations.
This is equivalent to the peak of the strain norm integrated over the celestial sphere because the angular harmonics are orthonormal.

For each sample in the \NRSur reference posterior, we use Eq.~\eqref{eq:peak-time} to infer the arrival time of the peak strain at geocenter, or any given detector.
This results in a posterior on the merger time at each detector. %
For the \NRSur run in the main text, the measured GPS peak times are $t_{\rm peak}^{\rm LHO} = \tPeakH$ at LIGO Hanford and $t_{\rm peak}^{\rm LLO} = \tPeakL$ at LIGO Livingston.
We select the maximum-likelihood sample from the preliminary reference posterior as a representative value for the ringdown analysis.
This is the reference time $\tRefSymbol = \tRef\, {\rm s}$ at geocenter, with the corresponding sky location $(\alpha = \raRef,\, \delta = \decRef)$ implicitly encoding the individual detector times.
We list all reference values in Table~\ref{tab:reference-parameters}.

To guide the analysis and report results, it is useful to define a reference timescale in units of the final black hole mass.
We derive a posterior on the redshifted final black hole mass $(1+z)\mf$ and spin $\chif$ by applying the \textsc{NRSur7dq4Remnant} model \cite{Varma:2019csw} to \NRSur posterior samples.
As reference to quote timescales in the ringdown analysis, we adopt the same maximum-likelihood sample mentioned above, which is quoted in Table~\ref{tab:reference-parameters} together with the corresponding timescale $\tMf$.
Relative to the chosen reference times, the peak distributions inferred by the \NRSur analysis in the main text are $t_{\rm peak}^{\rm LHO} - t_{\rm ref}^{\rm LHO} = \tPeakHrel$ at LIGO Hanford and $t_{\rm peak}^{\rm LLO} - t_{\rm ref}^{\rm LLO} = \tPeakLrel$ at LIGO Livingston.
The standard deviations of those distributions are $\tPeakHrelStd$ and $\tPeakLrelStd$ respectively.

\begin{table}
    \centering
    \begin{minipage}{0.7\textwidth}
    \centering
    \caption{Reference parameters for the ringdown analysis.}
    \label{tab:reference-parameters}
    \begin{ruledtabular}
    \begin{tabular}{lll}
        Parameter & Value & Description \\
        \midrule
        \tRefSymbol (geocenter) & \tRef~s & Reference time at geocenter \\
        $\alpha$ & \raRef~rad & Right ascension of source \\
        $\delta$ & \decRef~rad & Declination of source \\
        $\psi$ & \psiRef~rad & Polarization angle \\
        \midrule
        $t_{\rm peak}^{\rm LHO}$ & \tRefH~s & Reference time at LIGO Hanford \\
        $t_{\rm peak}^{\rm LLO}$ & \tRefL~s & Reference time at LIGO Livingston \\
        \midrule
        $\left(1+z\right)\mf$ & \mfRef~\Msun & Reference final black hole mass \\
        \tMf & \tMfRef & Reference $(1+z) \mf$ in units of time \\
        \midrule
        $\left(1+z\right) M$ & \mRef & Reference total mass \\
        $t_M$ & \tMRef & Reference $(1+z)M$ in units of time \\
    \end{tabular}
    \end{ruledtabular}
    \end{minipage}
\end{table}

\subsection{IIB. Ringdown analyses}

\subsubsection{IIB1. Parameterization and priors}

Each quasinormal mode is described by four parameters besides its frequency and damping rate: an overall amplitude $A_{\ell m n}$, a polarization ellipticity $\epsilon_{\ell m n}$, a polarization angle $\theta_{\ell m n}$, and a fiducial phase $\phi_{\ell m n}$ \cite{Isi:2021iql,Isi:2022mbx}.
These parameters control the amplitude and phase of the two gravitational-wave polarizations ($+$ and $\times$) for each mode:
\begin{subequations} \label{eq:hpc-ae}
    \begin{align}
        h_{+} &= A \left[ \cos\theta \cos (2\pi f t - \phi) - \epsilon \sin\theta \sin(2\pi f t - \phi)\right] \exp(-\gamma t) \, , \\
        h_{\times} &= A \left[ \sin\theta \cos (2\pi f t - \phi) + \epsilon \cos\theta \sin(2\pi f t - \phi)\right] \exp(-\gamma t) \, ,
    \end{align}
\end{subequations}
suppressing mode indices $(\ell, |m|, n)$ for brevity.
This is the most generic expression for a quasinormal-mode signal and subsumes both positive and negative frequency contributions, which jointly encode the polarization content of each mode \cite{Isi:2021iql,Isi:2022mbx}.
Assuming prograde modes with nonvanishing $m$, the sign of the frequency is related to the sign of the azimuthal number by $\mathrm{sgn}(m) = \mathrm{sgn} (f)$, so that the positive and negative frequencies encode the right- and left-handed polarized components of the mode.
The same would be true for retrograde modes, except that $\mathrm{sgn}(m) = - \mathrm{sgn} (f)$.
This paper only considers prograde modes.

Without information about the expected intrinsic amplitudes of the quasinormal modes, ringdown analyses cannot infer a luminosity distance and the only mass scale to which they are sensitive is the product $(1+z)\mf$.
In the case of a Kerr fit, all frequencies and damping rates are derived from a given $(1+z)\mf$ and $\chif$, such that $f_{\ell |m| n} = f_{\ell |m| n}[(1+z)\mf, \chif]$ and $\gamma_{\ell |m| n} = \gamma_{\ell |m| n}[(1+z)\mf, \chif]$.
For beyond-Kerr fits, we introduce additional parameters $\delta f_{\ell |m| n}$ and $\delta \gamma_{\ell |m| n}$ such that $f_{\ell |m| n} = f_{\ell |m| n}[(1+z)\mf, \chif]\exp(\delta f_{\ell |m| n})$ and $\gamma_{\ell |m| n} = \gamma_{\ell |m| n}[(1+z)\mf, \chif]\exp(\delta \gamma_{\ell |m| n})$.
The exponential parameterization avoids a singularity as the deviation parameter approaches $-1$ \cite{Isi:2021iql}.

The \ringdown code places priors that are flat in $(1+z)\mf$, $\chif$, $A$, $\theta$, and $\phi$ as well as, when applicable, $\delta f_{221}$ and $\delta \gamma_{221}$; the ellipticity prior peaks at $\epsilon = 0$ but has broad support over the entire domain (Fig.~16 in \cite{Isi:2022mbx}).
The amplitude prior is flat in $A$ over the interval $[0, 5\times 10^{-20}]$ and is broad enough to always offer full support to the posterior without truncating it.
The priors in $\delta f_{221}$ and $\delta \gamma_{221}$ are flat over the intervals $[-0.8, 0.8]$ and $[-0.5, 0.5]$ respectively.
The quasinormal-mode polarization angle $\theta$ is fully degenerate with the source polarization angle $\psi$ \cite{Isi:2021iql,Isi:2022mbx}, which defines the orientation of the source's angular momentum relative to the celestial North pole and is used to compute the antenna pattern response functions for each detector \cite{Anderson:T010110,Finn:2008np}.
Given this exact degeneracy, the \ringdown analysis chooses a fiducial angle $\psi = \psiRef$ to compute antenna patterns, based on the same maximum-likelihood reference sample used to derive $t_{\rm peak}$ as explained above, although any arbitrary choice of $\psi$ would be valid.
The prior in the \ringdown code is independent for each mode in all parameters.

The \pyring code natively parameterizes the modes slightly differently. For a given $(\ell, |m|, n)$ mode, it uses
\begin{align}
    h_+ - i h_\times =~  &C_{\ell, +m, n}\, {}_{-2}Y_{\ell +m}(\iota, \varphi=0) \exp\left[i\left(2\pi f_{\ell |m| n} t + \phi_{\ell, +m, n}\right)\right]  \exp\left[-\gamma_{\ell |m| n} t\right] + \nonumber \\
     &C_{\ell, -m, n}\, {}_{-2}Y_{\ell -m}(\iota, \varphi=0) \exp\left[i\left(-2\pi f_{\ell |m| n} t + \phi_{\ell, -m, n}\right)\right] \exp\left[-\gamma_{\ell |m| n} t\right] \, ,
\end{align}
where $\iota$ is an inclination parameter that is sampled from a prior uniform in $\cos\iota$, at the same time as the free amplitudes $C_{\ell, +m, n}$ and $C_{\ell, -m, n}$, which can be interpreted as the amplitude of the right and left-handed polarized contributions to the mode respectively; the corresponding phases, $\phi_{\ell, +m, n}$ and $\phi_{\ell, -m, n}$, combine in difference and sum to produce $\theta$ and $\phi$ in Eq.~\eqref{eq:hpc-ae} \cite{Isi:2022mbx}, and are fully degenerate with the polarization angle $\psi$, which \pyring also samples over a flat prior.
The \pyring prior is also flat on $(1+z)\mf$ and $\chif$.

The \pyring prior on the mode amplitude $A$ is implicitly defined by the priors on $C_{\ell,\pm m,n}$, which are uniform over the interval $[0, 5 \times 10^{-20}]$, and the prior on $\iota$; because of prior volume effects, it amounts to a density that has no support at the origin and disfavors $A \to 0$ (cf. Fig.~16 in \cite{Isi:2022mbx}).
Additionally, since all modes share $\cos\iota$, the prior correlates the different modes.
For a two-mode model with $\ell = |m| =2$ and $n=0,1$, the Jacobian to a flat prior in $\{A, \epsilon\}$ can be computed analytically and is given by
\begin{equation}
J = \frac{(1-\cos^2\iota)^{4}}{4\left[C_{2,-2,0}\left(1-\cos\iota\right)^{2}+C_{2,+2,0}\left(1+\cos\iota\right)^{2}\right] \left[C_{2,-2,1}\left(1-\cos\iota\right)^{2}+C_{2,+2,1}\left(1+\cos\iota\right)^{2}\right]} \,.
\end{equation}
In the main text, we reweight the \pyring posterior to a flat prior in $\{A, \epsilon\}$ by applying this Jacobian and truncating to the appropriate bounds.

The \pyring code implements deviations from the Kerr spectrum by writing
\begin{subequations}
\begin{align}
\omega_{221} &= \omega_{221} [(1+z)\mf, \chif] \left(1 + \delta \omega_{221}\right) \,,\\
\tau_{221} &= \tau_{221} [(1+z)\mf, \chif] \left(1 + \delta\tau_{221}\right) \, ,
\end{align}
\end{subequations}
where $\omega_{221} \equiv 2\pi f_{221}$ and $\tau_{221} \equiv 1/\gamma_{221}$.
For small deviation parameters, this parameterization is equivalent to the \ringdown one, except with a singularity at $\delta \omega = -1$ and $\delta \tau = -1$.
The above implies the following relationship between the \ringdown parameters $(\delta f_{221}, \delta \gamma_{221})$ and the \pyring parameters $(\delta \omega_{221}, \delta \tau_{221})$:
\begin{align}
\delta f_{221} = \log (1+\delta\omega_{221}) ~~, ~~ \delta \gamma_{221} = -\log (1+\delta\tau_{221})\, .
\end{align}
To go from a uniform prior in $(\delta \omega_{221}, \delta \tau_{221})$ to a uniform prior in $(\delta f_{221}, \delta \gamma_{221})$ we must apply a Jacobian given by
\begin{align}
J = \left| 1 + \delta \omega_{221} \right|^{-1} \left| 1 + \delta \tau_{221} \right|^{-1} .
\end{align}
In the main text, we reweight the \pyring posterior to a flat prior in $\delta f_{221}$ and $\delta \gamma_{221}$ by applying this Jacobian and truncating to the appropriate bounds.

\subsubsection{IIB2. Data conditioning}

Both the \pyring and \ringdown analyses are based on data sampled at 4096\,Hz with a covariance matrix derived from the same estimate of the power spectral density used in the main analysis described in Sec.~I of this Supplement.
Before obtaining the covariance matrix, the power spectral density is treated to censor frequencies below 20\,Hz and above 1830\,Hz to match the integration band for the likelihood in the full-signal analysis~\cite{Siegel:2023lxl}.

Additionally, \pyring applies a Butterworth bandpass filter to the data, suppressing frequencies below 20\,Hz and above 2043\,Hz; the filtering is applied to \PyRingSeglen\,s of data around the event.
\ringdown only applies a high-pass Butterworth filter at 10\,Hz to remove zero-frequency offsets; there is no low-pass filtering other than truncation of the frequency series at Nyquist (the digital filter described in Ref.~\cite{Siegel:2023lxl}); the conditioning is applied to 634\,s of data around the event time.
Neither code applies any filtering to the signal templates in the likelihood calculation~\cite{Siegel:2023lxl}.
The difference in conditioning at high frequencies between \ringdown and \pyring is understood to cause a subdominant (but measurable) systematic difference in the two posterior distributions.

Estimates of the SNR accumulated after the signal peak indicate that an integration time of $T = 0.6$\,s is sufficient to capture the entirety of the post-merger signal; therefore, this is the analysis duration used in all runs by the \ringdown code.
For reasons of computational efficiency, the \pyring analysis is run with a shorter integration time of $T = 0.2$\,s, leading to slightly broader posterior distributions.
Also to reduce computational cost, the \pyring analysis was run on a sparser grid of start times.

The main root of the systematic differences between the two codes is understood to be in the selection of the analysis data.
For each choice of analysis start time $\tStart$, the \ringdown code selects the first sample of the data to be analyzed at each detector based on the native sampling rate of 16384\,Hz provided by the LIGO detectors; once the sample closest to the requested start time is identified in the 16384\,Hz data, the \ringdown code downsamples to 4096\,Hz while preserving the selected sample.
The effective timing precision of \ringdown is thus $\delta t \approx 1/(16384\, {\rm Hz}) = 0.06 \, {\rm ms} \approx 0.18\, \tMf$.
The \pyring code, on the other hand, first downsamples the data to 4096\,Hz and then selects the first sample of the data to be analyzed at each detector, meaning that the effective timing precision is $\delta t \approx 1/(4096\, {\rm Hz}) = 0.24 \, {\rm ms} \approx 0.72\, \tMf$.
The effect of this coarse graining varies for each requested start time and for each detector: if the start time happens to fall on a sample at a given detector, it is unlikely to also fall on a sample at the other.

The above means that the \ringdown and \pyring results cannot be made to match by a uniform relabeling of the start times.
For example, for runs requesting $\tStart = \TimingDiscussionFundamentalTimeRingdown\,\tMf$ in the main text, the \ringdown code starts the analysis at $t_{\rm LHO} = \tGPSringdownFundamentalRefH\, {\rm s}$ (GPS) and $t_{\rm LLO} = \tGPSringdownFundamentalRefL\, {\rm s}$ (GPS) for Hanford and Livingston respectively, which is $\tGPSringdownFundamentalRefHdelta\, \tMf$ and $\tGPSringdownFundamentalRefLdelta\, \tMf$ relative to the requested \tStart in each detector respectively.
The closest available \pyring run is the one that requested $\tStart = \TimingDiscussionFundamentalTimePyring\, \tMf$; this starts the analysis at $t_{\rm LHO} = \tGPSpyringFundamentalRefH\, {\rm s}$ (GPS) and $t_{\rm LLO} = \tGPSpyringFundamentalRefL\, {\rm s}$ (GPS) for Hanford and Livingston respectively, which is $\tGPSpyringFundamentalRefHdelta\, \tMf$ and $\tGPSpyringFundamentalRefLdelta\, \tMf$ relative to the target time of $\tStart = \TimingDiscussionFundamentalTimeRingdown\,\tMf$ in each detector respectively.
In other words, the start time of the \pyring $\TimingDiscussionFundamentalTimePyring\, \tMf$ run differs from that in the \ringdown $\TimingDiscussionFundamentalTimeRingdown\, \tMf$ run by $\tGPSfundamentalPyringRingdownH\,\tMf$ for Hanford and $\tGPSfundamentalPyringRingdownL\, \tMf$ in Livingston.
A similar calculation shows that the \pyring $\TimingDiscussionOvertoneTimePyring\, \tMf$ start time is $\tGPSpyringOvertoneRefDelta\,\tMf$ after the \ringdown $\TimingDiscussionOvertoneTimeRingdown\, \tMf$ run in both detectors.
The fact that \pyring uses less data than \ringdown explains the systematic differences in the posteriors presented in the main text, as was verified by running \ringdown with the exact same data as \pyring and reproducing that code's results.

Just as the inspiral time-domain analysis, the two sets of post-merger analyses also ignore the uncertainty over the detector calibration, which is expected to have a negligible impact on ringdown analyses at this SNR~\cite{Sinha:2025snr}.

\subsubsection{IIB3. Computation of amplitude significance}

In the main text, we provide estimates for the significance with which we can
establish that the amplitude $A$ of a given quasinormal mode is greater than
zero. This entails estimating the posterior probability density at $A=0$, which
represents the boundary of the amplitude parameter space ($A \geq 0$) and
therefore will never be directly represented in the set of posterior samples.
Our significance estimates are based on the smallest probability $p$ such that
the highest posterior density (HPD) interval enclosing probability mass $p$
includes the origin.  HPD intervals can produce counter-intuitive significance
results when the posterior has a sharp truncation at large amplitude, either
because the prior cuts off the posterior or due to the structure of the
likelihood function, but our amplitude posteriors are not of this shape.  

We choose to estimate the significance of the amplitude $A > 0$ by direct
integration over an HPD interval of a KDE-based representation of the posterior
density, $p(A)$. This method has an advantage over sample-based methods in that
it can estimate arbitrarily small values of $p(0)$ (i.e.~arbitrarily high
significance), while direct sample-based methods bottom out at $p(0) \sim
1/N_\mathrm{samples}$. Compared to other, simpler estimates of the significance
such as calculating the $z$-score $z = \mu_A / \sigma_A$ where $\mu_A$ and
$\sigma_A$ are the mean and standard deviation of the posterior samples, this
method has the advantage that it can account for non-Gaussian shapes of the
posterior density.  We first form a standard KDE estimate of the posterior
density from the samples of the amplitude $A$, using automatic bandwidth
estimation from the \texttt{scipy.stats.gaussian\_kde} function.  This density
estimate, $k_\mathrm{raw}(A)$ is normalized such that it integrates to unity
over $-\infty < A < \infty$.  To account for the boundary at $A = 0$, we reflect
the density estimate about the origin, defining 
\begin{equation}
    k(A) = k_\mathrm{raw}(A) + k_\mathrm{raw}(-A);
\end{equation}
this ensures that $k(A)$ integrates to unity over $0 \leq A < \infty$, and so is
a suitable density estimate over our domain.

We then evaluate $k(0)$, the (estimated) posterior density at $A = 0$.  We
evaluate the significance of the amplitude $A > 0$ by integrating the posterior
density estimate, $k$, over $A$ values such that $k(A) > k(0)$, i.e., computing
the smallest $p$ such that the highest-posterior-density interval containing
probability mass $p$ includes $A = 0$.  Thus,
\begin{equation}
    p = \int_{\left\{ A \mid k(A) > k(0) \right\}} \mathrm{d} A \, k(A).
\end{equation}
By construction, $0 \leq p \leq 1$.  

Once we have an estimate for $p$, we communicate this as a number of $\sigma$ by
expressing this value in terms of tail probabilities for the Gaussian
distribution, i.e., we establish that $A>0$ at $x\,\sigma$ significance, where
$x$ is defined by
\begin{equation}
\int_{-x}^x \mathrm{d} x' \, \phi\left( x' \right) = p \, ,
\end{equation}
where $\phi$ is the probability density function for the standard normal distribution.  Equivalently (but more stably numerically), we can define $x$ by 
\begin{equation}
\int_{-\infty}^{-x} \mathrm{d} x' \, \phi\left( x' \right) = \frac{1 - p}{2} \, .
\end{equation}
This is the quantity that we estimate and report in the main text.  The quantity $1-p$ can be computed by 
\begin{equation}
    1-p = \int_{\left\{ A \mid k(A) \leq k(0) \right\}} \mathrm{d} A \, k(A),
\end{equation}
which is more numerically stable to evaluate when $p \simeq 1$.

\subsection{IIC. Further parameter posteriors}

Figure~\textcolor{red}{5} in the main text shows posteriors for the amplitude of the fundamental and overtone modes as a function of the analysis start time.
Amplitudes are referenced to the start time of each analysis and are presented for both the single-mode and two-mode models.
For completeness, Fig.~\ref{fig:supp_modes} shows the posteriors for the frequency and damping rate of each mode in similar style.
In the bottom panel, we further show the network matched-filtered SNR as a function of time for both models.
For the two-mode analysis, we present the SNR of the full model, as the mode nonorthogonality makes defining the SNR of each mode ambiguous.
The inferred frequencies and damping rates are consistent across all times of applicability of their respective model, i.e., after $\tSecondOvertone\,\tMf$ for the two-mode model.
Uncertainties increase and the SNR decreases as the analysis start time is moved later, as expected.
Moreover, the SNR recovered by the two-mode and the one-mode models in their overlapping time region are highly consistent.

\begin{figure}
    \centering
    \includegraphics[width=\linewidth]{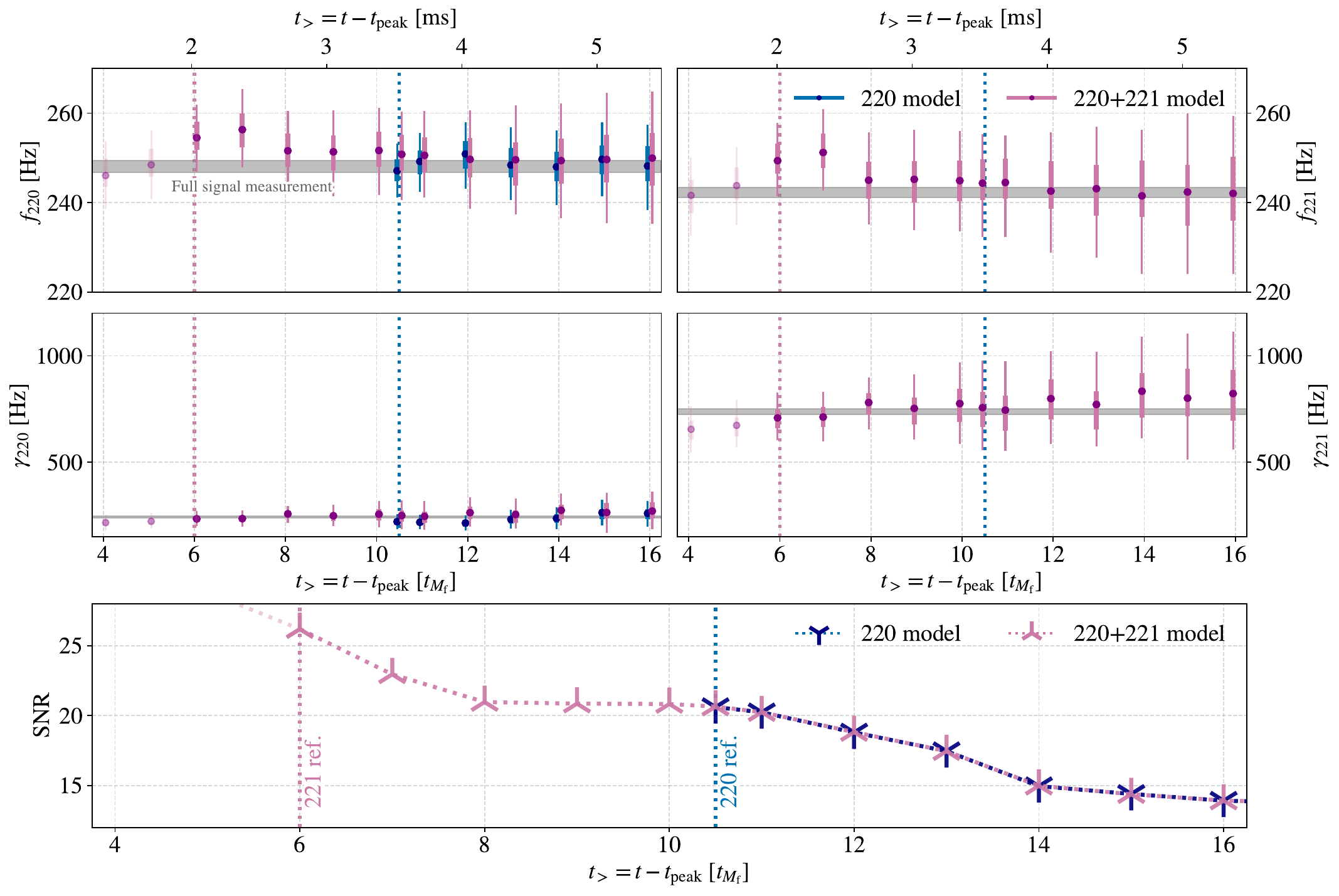}
    \caption{Similar to Fig.~\textcolor{red}{3} in the main text, but showing the frequency (left) and damping rate (right) of the fundamental (top) and overtone (middle) modes as a function of the analysis start time.
    The bottom panel shows the network matched-filtered SNR as a function of time for both the single-mode and two-mode models, which are highly consistent.
    In the two-mode model, both frequencies and damping rates are inferred jointly from a common mass and spin assuming a Kerr spectrum.
    We show the 90\% credible intervals for the frequencies and damping times inferred from the full signal analysis as grey horizontal shaded areas.
    As in Fig.~\textcolor{red}{3} in the main text, we show results with \ringdown as there is not exact timing correspondence between the \ringdown and \pyring posteriors; we nonetheless obtain similar results with \pyring.
    }
    \label{fig:supp_modes}
\end{figure}

\section{III. Technical details and further tests for the pre-merger analysis}

To enable the sharp truncation of the data and gravitational-wave model at a time before the binary merger~\cite{Isi:2020tac,Cabero:2017avf,Correia:2023ipz}, we carry out the analysis in the time-domain using \textsc{TDinf} \cite{Miller:2023ncs,Miller:2025eak}, guided by the reference parameters from Table~\ref{tab:reference-parameters}.
The \textsc{TDinf} inference package samples the 15-dimensional parameter space of a quasicircular black-hole binary using the \textsc{emcee} sampler \cite{ForemanMackey:2013}.
We adopt the same settings as the full-signal frequency-domain analyses with \bilby (in terms of the data, trigger time, power spectral density, and bandwidth), other than the amount of data considered, certain priors, and the treatment of calibration.
Time-domain analysis of the full signal is based on \TDinfSeglen\,s of data, which are appropriately truncated.
The priors are the same as the \bilby analysis other than the masses (uniform in total mass and mass ratio), distance (uniform in luminosity distance), and time (Gaussian centered around the geocenter trigger time with a standard deviation of \TDinfTimeStd\,s).
The distance prior has a minimal effect on the inferred area, which is based on redshifted masses that are minimally correlated with the distance. 
The composite area law prior is broad and relatively flat, as seen in Fig.~\textcolor{red}{5} in the main text.
We use the autocorrelation length (ACL) from the full ensemble of walkers to determine the burn-in period and thinning of the chains. 
Depending on the truncation time, the lengths of these chains range between 100,000--500,000 steps, each with 512 walkers.
We use a burn-in of at least five times the maximum ACL across sampled parameters, corresponding to $>40\%$ the chain length, and thin by half of the minimum ACL. 
For some of the truncation times, a handful of chains are not converged to the bulk of the posterior at the end of the burn-in, instead representing a secondary mode in likelihood and sky location.
For these chains, we manually extend the burn-in period until they converge to the bulk of the posterior.
We have verified that the posterior is not restricted by any of the prior edges.
We have also verified that the time-domain and \bilby analyses of the full signal yield statistically identical posteriors when reweighted to the same prior for both \NRSur and \Phenom.

An additional difference is that the time-domain analysis neglects the uncertainty over the detector calibration, which \bilby marginalizes over.
We have verified that this has a minimal impact by repeating the \bilby analysis while neglecting calibration uncertainty.
The posteriors for the detector-frame total mass with and without marginalizing over the calibration uncertainty differ by a Jensen\textendash Shannon divergence~\cite{lin:1991} of \JSNRSurCalibration\,nat.
For reference, 
changing the waveform model from \NRSur to \Phenom yields a Jensen\textendash Shannon divergence of \JSNRSurvsPhenom\,nat, to \EOB of \JSNRSurvsEOB\,nat, and to \PhenomOfour of \JSNRSurvsPhenomOfour\,nat.
The impact of calibration uncertainty is therefore subdominant compared to other potential sources of systematic errors. 

The main text presents results for a number of truncation times in Fig.~\textcolor{red}{5}.
We use the gravitational-wave luminosity inferred from the full-signal analysis (a measure of how dynamical and relativistic the system is~\cite{Blanchet:2013haa}) to determine that most of the emission occurs during the late coalescence stages.
While the averaged gravitational-wave strain, defined as $\left(\sum_{\ell,m}|h_{\ell m}(t)|^2\right)^{1/2}$, peaks at $\tRefSymbol$ by definition, the flux, proportional to $\left(\sum_{\ell,m}|\dot{h}_{\ell m}(t)|^2\right)^{1/2}$ (where a dot denotes differentiation with respect to time), peaks at $\FluxPeak\,\tMtot$ after $\tRefSymbol$. 
Specifically, the flux is 10\% of its maximum $\FluxTenPercent\,\tMtot$ before $\tRefSymbol$ and 1\% of its maximum $\FluxOnePercent\,\tMtot$ before $\tRefSymbol$.
The former is also comparable to common estimates for the transition to merger~\cite{PhysRevD.50.5025,Baumgarte:2000mk,Blanchet:2025agj}.
Based on these time estimates, Fig.~\textcolor{red}{5} in the main text presents results for an array of truncation times after $\tEnd = -250\,\tMtot$ and 
more detailed results for $\tEnd=\Pretime\,\tMtot$.

\section{IV. Technical details and further results for the area law analysis}

The area law test relies on the independent analyses of the pre-merger data described in Sec.~III and the post-merger data described in Sec.~II of this Supplement. 
Both tests are carried out in the time-domain and avoid quantitative reference to the full-signal results, which are based on waveform models within general relativity that obey the area law by construction.
Since the pre- and post-merger analyses are also independent of each other, there is no information shared between the two sets of results and no assumption about the initial and final black holes sharing the same location in the sky.
The only slight exception to this is that, by comparing initial and final quantities in the detector frame, we implicitly assume that the pre- and post-merger signals were redshifted by the same factor.
The recoil kick of the remnant can break this assumption by inducing a redshift; however, this is at most $10\%$ and only in the most extreme cases with large spins inconsistent with \eventname~\cite{Healy:2022jbh}.
An analysis in which the extrinsic parameters are sampled jointly for both the pre- and post-merger data, such as \cite{Correia:2023ipz}, could gain by imposing the same location in the sky for the initial and final black holes.

Our test yields independent measurements of the initial and final areas.
If the posterior distributions on the final and initial area are denoted as \(p_f(\mathcal{A}_{\rm f})\) and \(p_i(\mathcal{A}_{\rm i})\) respectively,
the significance \(\sigma\) of a non-detection of a violation of Hawking's area law is given by the separation between \(p_f(\mathcal{A}_{\rm f})\) and \(p_i(\mathcal{A}_{\rm i})\).
Following the cosmology literature estimating the significance of the Hubble tension~\cite{Raveri:2018wln,Verde:2019ivm}, we estimate the significance as $X \sigma$ with
\begin{equation}
    X =\frac{\mu_{\rm f}-\mu_{\rm i}}{\sqrt{\sigma_{\rm f}^2+\sigma_{\rm i}^2}}\,,
    \label{eq:sigmaG}
\end{equation}
where $\mu_{\rm i}/\mu_{\rm f}$ and $\sigma_{\rm i}/\sigma_{\rm f}$ are the means and standard deviations respectively of the initial and final area distributions.
Since this estimate relies only on the first two cumulants, it is less sensitive to sampling error at the tails of the distribution.
Empirically, sampling the posterior tails beyond the ${\sim}4\sigma$ level is unreliable and highly sensitive to sampler settings and minor analysis choices, so we avoid using tail samples.

In the main text, Fig.~\textcolor{red}{5} shows results where the final area has been measured via a single $(\ell=2,|m|=2,n=0)$ mode starting at $\tOvertone\,\tMf$, which is the earliest time that the overtone significance falls below $1\sigma$.
Identification of this time is based on the post-merger data alone, without requesting consistency with the full-signal analysis.
Here we show similar results obtained for different pre- and post-merger start times, for both the single-mode and two-mode ringdown models.
Figure~\ref{fig:supp_areas} displays the inferred initial and final areas as a function of inspiral end time $\tEnd$ and ringdown start time $\tStart$, respectively.
Similarly to Fig.~\textcolor{red}{5}, we show the 90\% credible intervals for the initial and final areas as inferred from the full-signal analysis for reference.
These are consistent with both the initial and final areas within statistical uncertainties, although this comparison does not guide any of the analysis choices in testing the area law.
The significance, as defined in Eq.~\eqref{eq:sigmaG}, can be assessed via the separation of the distributions relative to their widths.

As expected, the uncertainty in the remnant area grows as the start time is pushed to later parts of the data where the signal is weaker;
for the initial area, the opposite is true and the uncertainty grows as the analysis end time is pushed earlier.
The single-mode ringdown result used in the main text corresponds to a start time of $\tStart = \tOvertone\,\tMf$ and is highlighted by a vertical blue band.
Including the overtone in the model at this and later times necessarily broadens the uncertainty in the final area.
This is both because the overtone is not required to explain the data and because the overtone and the fundamental are not orthogonal: introducing additional, unconstrained degrees of freedom broadens the posterior distribution.

The reference start time showcased in the main text was chosen independently of the full-signal analysis purely as the earliest time at which the data are consistent with a single mode, according to our $1\sigma$ criterion.
Had the SNR of this event been higher, we would likely not have found the data to be sufficiently well-described by a single mode at $\tStart = \tOvertone\,\tMf$.
This would have led us to use a later start time for the reference ringdown analysis with a single mode.
Carrying out the area law test with the two-mode analysis instead selects $\tSecondOvertone\,\tMf$ as the earliest time at which this model explains the observed data.
The area law is again confidently satisfied at $\ALsignOvertone\sigma$ significance.

\begin{figure}
    \centering
    \includegraphics[width=\linewidth]{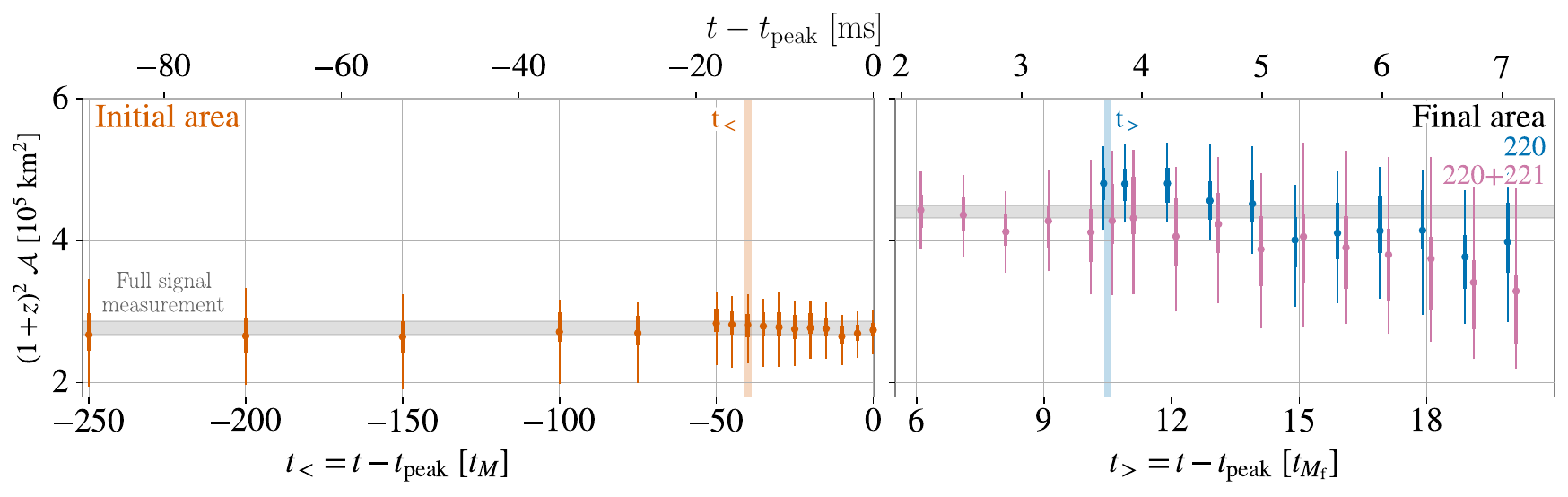}
    \caption{Measurements for the initial and final black hole areas as a function of fit time. 
    (Left panel) Initial black hole area inferred by analyzing the pre-merger signal with \textsc{TDinf}, truncated at the time indicated on the $x$-axis. 
    (Right panel) Final black hole areas inferred by analyzing the post-merger signal with \ringdown, with the analysis start time indicated on the $x$-axis.
    We show results for the two-mode analysis from $\tSecondOvertone\,\tMf$ after the peak strain and the single-mode only from $\tOvertone\,\tMf$ after the peak strain. 
    We highlight with vertical bands the reference areas quoted in this paper, at $\Pretime\,\tMtot$ before the peak strain for the inspiral analysis and $\tOvertone\,\tMf$ after the peak strain for the single-mode ringdown analysis. 
    Horizontal grey bands indicate the 90\% credible intervals for the initial and final areas as inferred from the full inspiral-merger-ringdown analysis.
    }
    \label{fig:supp_areas}
\end{figure}
 \fi

\end{document}